\documentclass[10pt]{article}
\pdfoutput=1
\usepackage[utf8]{inputenc}
\usepackage[T1]{fontenc}
\usepackage[pdftex]{geometry}
\usepackage[english]{babel}
\usepackage[authoryear]{natbib}
\usepackage{textcomp,amsmath,amssymb,pdflscape,longtable,afterpage,calc,booktabs}
\usepackage{url} 
\usepackage{graphicx}
\newcommand{\partdrv}[2]{\frac{\partial #1}{\partial #2}}  
\newcommand{\chem}[3][]{\ensuremath{\mathrm{{#2}}_{#3}^{#1}}}  
\newcommand{\ion}[2]{\ensuremath{\text{#1}^{#2}}}  

\begin{document}
\bibliographystyle{model2-names}
\let\WriteBookmarks\relax
\def\floatpagepagefraction{1}
\def\textpagefraction{.001}
\thispagestyle{empty}
\begin{center}
\Large Thomas Ruedas\textsuperscript{1,2}\\[2ex]
Doris Breuer\textsuperscript{2}\\[5ex]
\textbf{Electrical and seismological structure of the martian mantle and the detectability of impact-generated anomalies}\\[5ex]
final version\\[5ex]
18 September 2020\\[10ex]
published:\\
\textit{Icarus} 358, 114176 (2021)\\[15ex]
\normalsize
\textsuperscript{1}Museum f\"ur Naturkunde Berlin, Germany\\[5ex]
\textsuperscript{2}Institute of Planetary Research, German Aerospace Center (DLR), Berlin, Germany
\rule{0pt}{12pt}
\end{center}
\vfill
\footnotesize The version of record is available at \url{http://dx.doi.org/10.1016/j.icarus.2020.114176}.\\
This author pre-print version is shared under the Creative Commons Attribution Non-Commercial No Derivatives License (CC BY-NC-ND 4.0).
\normalsize

\title{Electrical and seismological structure of the martian mantle and the detectability of impact-generated anomalies}

\author{Thomas Ruedas\thanks{Corresponding author: T. Ruedas, Museum f\"ur Naturkunde Berlin, Germany (Thomas.Ruedas@mfn.berlin)}\\{\footnotesize Museum f\"ur Naturkunde Berlin, Germany}\\{\footnotesize Institute of Planetary Research, German Aerospace Center (DLR), Berlin, Germany}\\[2ex]
Doris Breuer\\{\footnotesize Institute of Planetary Research, German Aerospace Center (DLR), Berlin, Germany}}
\date{}
\maketitle

\textbf{Highlights}
\begin{itemize}
\item Geophysical subsurface impact signatures are detectable under favorable conditions.
\item A combination of several methods will be necessary for basin identification.
\item Electromagnetic methods are most promising for investigating water concentrations.
\item Signatures hold information about impact melt dynamics.
\end{itemize}

\begin{flushleft}
Mars, interior; Impact processes
\end{flushleft}

\begin{abstract}
We derive synthetic electrical conductivity, seismic velocity, and density distributions from the results of martian mantle convection models affected by basin-forming meteorite impacts. The electrical conductivity features an intermediate minimum in the strongly depleted topmost mantle, sandwiched between higher conductivities in the lower crust and a smooth increase toward almost constant high values at depths greater than 400\,km. The bulk sound speed increases mostly smoothly throughout the mantle, with only one marked change at the appearance of $\beta$-olivine near 1100\,km depth. An assessment of the detectability of the subsurface traces of an impact suggests that its signature would be visible in both observables at least if efficient melt extraction from the shock-molten target occurs, but it will not always be particularly conspicuous even for large basins; observations with extensive spatial and temporal coverage would improve their detectability. Electromagnetic sounding may offer another possibility to investigate the properties of the mantle, especially in regions of impact structures. In comparison with seismology and gravimetry, its application to the martian interior has received little consideration so far. Of particular interest is its potential for constraining the water content of the mantle. By comparing electromagnetic sounding data of an impact structure with model predictions, it might also be possible to answer the open question of the efficiency of extraction of impact-generated melt.
\end{abstract}

\section{Introduction}
The electrical, seismic, and density structures of the mantles of terrestrial planets reflect in multiple ways the physical and chemical state and history of their interiors, but observations of any of them are only snapshots of the present situation and lack the temporal dimension. By contrast, dynamical numerical models of mantle evolution provide both spatial and temporal information and can therefore give insights into processes of whose products we can today only observe the scrambled and diluted remains with often ambiguous signatures. Convection models per se do not need information about seismic and electrical properties as input, and many of them use only some fairly simple representation of the density, and so they are generally not suited for deriving the seismic or electrical structure. Nonetheless, it is possible to integrate a more detailed description of the mantle's minerals and their properties into them and thus derive models of seismic and electrical structure from the same mineralogical model as the dynamical evolution model.\par
The prediction of seismic velocities and/or electrical conductivity from such convection models has already been attempted in some previous studies of terrestrial mantle convection but to our knowledge not for other planets. \citet{Goes:etal04} ran a suite of extended Boussinesq models of thermal convection that included phase transitions and a depth dependence of some thermoelastic properties. The authors used some material parameters adjusted to be compatible with the compilation of seismic property data that they then used to derive seismic velocities and attenuation models from the dynamical models. However, they did not include compositional changes due to melting, and the thermoelastic property framework was not internally self-consistent. \citet{Kreu:etal04} and \citet{Ruedas06} used a similar type of convection models that did include melting and its effect on composition but derived seismic velocity anomalies (rather than absolute velocities) and attenuation from empirical rock-physical relations. They also calculated an electrical conductivity model based on limited experimental data for some minerals, whole rock, and silicate melts, whereby the latter study also included the effect of water. \citet{Naka:etal09} and \citet{Tiro:etal12} were the first to combine dynamical convection models with internally self-consistent thermoelastic models, which makes the calculation of density variations an integral part of the dynamic model, and used them to predict seismic observables. However, both studies limited their application only to S-wave anomalies, although P-wave distributions would have been a straightforward by-product of their models. In the present study, we combine information from an internally consistent set of mineral end-member thermoelastic properties with an empirical model of mantle petrology to calculate the material properties of the martian mantle and crust. These properties are then used in both the dynamical model and the thermoelastic observables that yield a model of the density and seismological structure of the silicate part of Mars. The same petrological model is also used for the derivation of electrical conductivity distributions. As a major focus of this paper lies on processes in which an impact is large enough to affect the mantle and to trigger a longer-term regional effect, the emphasis of our considerations lies on large-scale effects on the mantle and crust rather than on local structure.\par
Some of the oldest, largest, and in some cases most conspicuous features of the martian surface are basins formed by large impacts. The largest of these basins, such as Utopia, Ares, or Acidalia, have diameters of more than 3000\,km, and the impacts thought to have produced them must have had substantial regional or even global effects on the martian interior, because they have penetrated down to the sublithospheric mantle and exerted a direct influence on convection and melting in the mantle \citep[e.g.,][]{Rees:etal02,Rees:etal04,WAWatt:etal09,JHRobe:etal09,Gola:etal11,JHRoAr-Ha12,RuBr17c}. \citet{RuBr17c} showed that the thermal and chemical anomaly produced by the impact and concomitant melting is deformed and successively obliterated by the vigorous convective mantle currents it has triggered. However, under assumptions that increase the viscosity of the mantle, e.g., a low water content, remnants of the chemical anomaly can survive to the present. These anomalies should in principle also be detectable by geophysical methods, but such methods have so far been restricted to orbiting spacecraft, most importantly gravity and magnetic observations. There is more than one possible interpretation of such potential field signatures, however, and further assumptions have to be made to constrain them. Observations of Mars' magnetic field provide information mostly on crustal magnetization, and processes like shock or thermal demagnetization or re-magnetization of target materials in the presence of an external field \citep[e.g.,][]{Lill:etal13a} make them more difficult to interpret.\par
In this paper, we revisit and supplement some of the models from \citet{RuBr17c}, which tracked the evolution of a Mars-like planet that experienced a single basin-forming impact, and calculate the resulting anomalies of some geophysical observables. Our earlier study was mostly concerned with various aspects of the dynamical evolution and its relations with heat flow as well as with gravity. The focus in the present paper lies mostly on assessing the necessity of more ground-based geophysical studies, i.e., on methods that can not or only with limitations be applied from orbiters but potentially deliver less ambiguous results than gravity or magnetic measurements on both global structure and regional features such as impact sites. These methods are especially electromagnetic and seismic soundings. The latter is being pioneered now by the InSight mission \citep[e.g.,][]{Logn:etal19,Pann:etal17}, but the former has received little attention so far. There have been some electromagnetic soundings of the Earth's mantle using magnetometer data of orbiting satellites \citep[e.g.,][]{CoCo04}, but the only evaluation of measurements at Mars known to us is an attempt by \citet{CiTa14}, who derived a radial conductivity profile of the martian mantle from Mars Global Surveyor (MGS) magnetometer data. It may be possible to derive usable data from measurements of the magnetometer on the InSight lander as a by-product of that mission \citep[e.g.,][]{PJChi:etal19,CTRuss:etal19,AMitt:etal20a,YYu:etal20}, although those data are not expected to be of the highest quality and will be affected by noise from the lander. Nonetheless both methods have the potential to elucidate the large-scale basic structure of the martian mantle and constrain dynamically relevant properties such as the crustal thickness or the position and width of phase boundaries.\par
The motivation for applying electromagnetic and seismic methods to impact structures in addition to the data already available is twofold. First, there are cases of basins with diameters above 1000\,km, especially among very early structures, whose identity as being impact-generated is not firmly established.  For instance, the impact origin of Chryse Planitia, which has a diameter of 1725\,km and an age similar to the aforementioned three largest basins, has long been disputed and is still being studied \citep[e.g.,][]{LPan:etal19}. The list of impact basins by \citet{Frey08}, which has been used in several dynamical studies, contains 20 entries, and follow-up work by \citet{FrMa14} ramps this count up to as many as 31 candidate structures, which are rated according to criteria such as topographic or crustal thickness expression. On the other hand, \citet{BoAn-Ha17} argued that the observation that the dichotomy boundary is crossed by only one basin of this magnitude (Isidis) makes a number of more than 12 basin-forming impacts between the dichotomy-forming event and the final cadence defined by Hellas, Utopia, Isidis, and Argyre highly unlikely from a statistical point of view. If additional means such as geophysical signatures originating in the deeper interior are available to confirm or dismiss candidate impact structures, the impact count and in consequence the influence of impacts on martian evolution could be better assessed.\par
The second incentive is to help quantify how impact-generated melt is extracted from its source after the impact. The extreme heating by the shock is expected to melt a part of the target, and melting efficiency increases if the target was already close to the solidus temperature before the impact, as would be the case for a very large event that reaches into the mantle. Factors such as the high permeability of the partially molten target or the dynamics of post-impact uplift can be imagined to promote massive eruption of impact melts. In this case, the residue would be very strongly depleted in fusible components and be even more stripped of incompatible trace components than the background mantle, and a thick post-impact crust with an abnormally low concentration in heat-producing elements and volatiles is expected to form. On the other hand, it is also conceivable that at least a part of the melt remains trapped in the interior and recrystallizes, because the lifetime of potential melt migration pathways after the impact is too short or the prevailing stress field is unfavorable to melt extraction. As a consequence, the target mantle would be much less or not at all depleted, and the pre-impact crust blasted off by the impact would not or only partly be replaced by erupting post-impact melt. This would potentially explain the reduced crustal thicknesses inferred from gravity for various large basins such as Hellas on Mars \citep[e.g.,][]{GANeum:etal04} or the South Pole--Aitken basin on the Moon \citep[e.g.,][]{Wiec:etal13}. The presence or absence of geophysical signatures that depend on the efficiency of impact melt extraction in confirmed impact structures could therefore put constraints on models of melt production during the impact process. Their dynamics are only beginning to be investigated in detail \citep[e.g.,][]{Mans:etal19a,Mans:etal21}, and they could also improve the treatment of impacts in evolutionary geodynamical models. The fate of this melt is likely to have implications for the local dynamics of the mantle and its melt productivity in the post-impact phase and for the resulting geochemical variability. In the literature, scenarios ranging from no extraction \citep{Pado:etal17} to substantial extraction \citep{Gola:etal11,RuBr17c} have been considered. By using models from the latter study as the starting point for this investigation, we assume a scenario that would maximize the expected effects.  However, in order to assess the importance of extraction, we supplemented those models with some additional ones in which impact melt extraction was suppressed.

\section{Method}
The models of \citet{RuBr17c} were fully dynamical mantle convection calculations on a two-dimensional spherical annulus grid with pressure-, temperature-, and composition-dependent viscosity and internal and basal heating, carried out with a modified version of StagYY \citep{Tackley96a,Tackley08,HeTa08} in the anelastic, compressible approximation; a list of specific model parameters and the characteristics of the impact basins is provided in Table~\ref{tab:modpar}. They were coupled with a detailed mineral physics model of martian mineralogy after \citet{BeFe97} based on the bulk silicate Mars composition from \citet{WaDr94}. That model combines thermoelastic data for the mineral end-members with experiment-based parameterizations of the phase diagram and petrology of the martian mantle and empirical formulas for various major- and minor-element abundances in the mineral solid solutions, in particular for Mg--Fe partitioning. The partitioning formulas for the individual mineral phases were taken from the literature or constructed from available published experimental data. They describe the dependencies of the partitioning coefficients on pressure, temperature, and composition to the (quite variable) extent permitted by the data; full details can be found in the Supporting Information of \citet{RuBr17c}. The model therefore reproduces closely the martian mantle petrology for the chosen bulk silicate composition and ensures that the same physical and chemical properties that controlled the dynamical evolution of those convection models are now also used for the derivation of seismic velocities and electrical conductivity. As we do not have a similar representation for crustal material, the crust was assigned a homogeneous MORB-like mineralogy consisting mostly of clinopyroxene and plagioclase \citep[e.g.,][]{LiOh07}, although its water content is derived from the dynamical model. Compositional variability such as lithological inhomogeneities or depletion due to melting is tracked with tracer particles.  The melt dynamics is simplified by extracting segregating melt instantaneously and adding it to the top of the corresponding column of the numerical grid.\par
Melting was included in the dynamical models as a cause of chemical variations that affected the dynamical behavior and also left its signature in geophysical observables, even though no melt is found to exist in the martian interior today on a global scale.  This is consistent with other models of the thermal evolution of Mars \citep[e.g.,][]{Ples:etal18b}, which also find the conditions for melting to be fulfilled at most at some isolated local sites in modern Mars. The key compositional features considered in connection with melting were the iron, radioactive element, and water concentrations. The latter was varied between the nominal bulk silicate Mars value from \citet{WaDr94} of 36\,ppm and its two- and fourfold (72 and 144\,ppm) with the purpose of probing its effect on convective vigor.\par
An impact is imposed as an instantaneous thermal anomaly with a shape determined from simple scaling laws and a temperature distribution derived from an impedance match calculation and empirical fits to the shock pressure decay from numerical impact simulations \citep{Rees:etal02,WAWatt:etal09,Ruedas17a}. The impactor strikes at 4\,Ga, which is 400\,My after the start of the modeled evolution, and is specified as having a magnitude that would result in a basin of the size of either Huygens, Isidis, or Utopia. Each of them stands as a representative example of a size class: the Huygens impact produced a crater a few hundred kilometers wide and affected mostly the lithosphere; the Isidis impact formed a basin with a diameter of $\sim$1000\,km and had effects that reached below the lithosphere; the Utopia impact resulted in a basin of almost 3400\,km and melted even parts of the mantle below the maximum depth of regular melting at its time of formation (see Table~\ref{tab:modpar}). The site of the impact is chosen such that a coincidence with a hot upwelling or a cold downwelling is avoided as far as possible in an attempt to produce an effect representative for the mean thermal structure of the target, as local variations would modify the thermal effects of shock-heating; therefore, the impacts are at different locations in models with different water contents.\par
Almost all of the melt produced by it and in its aftermath is extracted to build the crust and is assumed to fill the near-surface pore space that is imposed at the top of the crust, whereby it is assumed to lose heat very efficiently and thus attain surface temperature immediately, i.e., to crystallize within a single timestep, which is typically on the order of 10\,ky.  According to independent estimates \citep[e.g.,][]{ReSo06,Rees:etal07a}, the crystallization of the impact-generated melt pond happens to occur on similar or even shorter timescales. More importantly, the crust forming by the crystallization of the post-impact melt pool will have a very low porosity. In this paper, we refer to these processes, which result in the impact basin being turned into a low-porosity anomaly, as ``porosity deletion'' or similar terms. As the model algorithm cannot handle a completely liquid body inside the essentially solid planet, a maximum melting degree that can be reached has to be set and is fixed here at 60\%, which corresponds to the absence of practically all fusible components (pyroxenes, most of the Al-bearing phases and fayalite fraction in olivine). Such upper limits to the contribution of melts to crust formation have also been defined in other studies of impact--convection interactions \citep[e.g.,][]{Gola:etal11}. They effectively entail an upper bound on the thickness of the post-impact melt sheet and crust and imply immediate in-situ recrystallization in, or inefficient melt extraction from, the central parts of the shocked source region to the surface for some part of the target. We expect that without this restriction, the impact melt would be even more mobile, and the potentially extracted melt volume and the magma pond it forms would be even larger and have a more strongly ultramafic composition. The observed seismic and electric anomalies in the mantle source regions of these melts should also be more pronounced in this case. When the magma pond crystallizes and differentiates, melts from source regions with different melting degrees are presumably mixed, and the mineralogy that results after crystallization represents an average composition of all melts, such that the geochemical signal from the most extreme melting is muted. Some implications of the efficiency of melt extraction for the formation of post-impact crust will be discussed in Sect.~\ref{sect:meltextr}.\par
We use a pressure-dependent porosity profile to describe the decay of porosity with depth in the megaregolith and its effects on density and thermal conductivity. The result of porosity deletion by impact melts is then the erasure of that porosity; we do not model the temporal changes in porosity due to impacts of smaller sizes throughout the history of the planet, and hence, porosity once destroyed by melt is not restored. By doing so, we ignore the new formation of regolith by the constant flux of smaller meteorites, but on the length-scale of megaregolith that reaches to several kilometers depth, this assumption should be acceptable, because at this stage in planetary evolution no substantial new formation of megaregolith is expected, and it is only this large length-scale that can be resolved at least rudimentarily in our global model at all. Recent analyses of the shallow density structure in and around large lunar impact craters determined that the porosity in many of them is indeed substantially reduced, probably as a consequence of impact melt filling the pore space \citep[and D.~Wahl, pers. comm.]{WaOb19,Wahl:etal20}. This simplified treatment facilitates the formation of crustal density anomalies and has implications for the geophysical signatures of impacts, as will be discussed below. Furthermore, variants of the models with the lowest water content were run in which the extraction of melt generated directly by the impact was suppressed. As a consequence, that melt was assumed to recrystallize in situ in those cases, and magma-flooding of the crater and the concomitant deletion of porosity did not occur. Melt produced by magmatism in the millions of years following the impact is extracted in both scenarios. In this paper, our attention will mostly be directed at the models with impact melt extraction, however. Finally, an additional impact-free model was also included for each of the three initial water concentrations considered; these models were called ``reference models'' in \citet{RuBr17c}.\par
\begin{table}
\caption{General model parameters.\label{tab:modpar}}
\begin{tabular*}{\textwidth}{@{}lcccl@{}}\toprule
Planetary radius, $R_\mathrm{P}$&\multicolumn{3}{c}{3389.5\,km}&\citet{Arch:etal11}\\
Total planetary mass, $M$&\multicolumn{3}{c}{$6.4185\cdot10^{23}$\,kg}&\citet{Kono:etal11,Jacobson10}\\
Surface temperature&\multicolumn{3}{c}{218\,K}&\citet{Catling15}\\
\textit{Mantle}\\
Mantle thickness, $z_\mathrm{m}$&\multicolumn{3}{c}{1659.5\,km}&\citet{Rivo:etal11}\\
Initial potential temperature, $T_\mathrm{pot}$&\multicolumn{3}{c}{1700\,K}&\\
Initial core superheating&\multicolumn{3}{c}{150\,K}&\\
Surface porosity, $\varphi_\mathrm{surf}$&\multicolumn{3}{c}{0.2}&\citet{Clifford93}\\
Melt extraction threshold, $\varphi_\mathrm{r}$&\multicolumn{3}{c}{0.007}&after \citet{Faul01}\\
Grain size, $d$&\multicolumn{3}{c}{1\,cm}&\citet{HiKo03}\\
Bulk silicate Mars Mg\#&\multicolumn{3}{c}{0.75}&\citet{WaDr94}\\
Present-day K content&\multicolumn{3}{c}{305\,ppm}&\citet{WaDr94}\\
Present-day Th content&\multicolumn{3}{c}{56\,ppb}&\citet{WaDr94}\\
Present-day U content&\multicolumn{3}{c}{16\,ppb}&\citet{WaDr94}\\
Initial bulk water content&\multicolumn{3}{c}{36, 72, 144\,ppm}&after \citet{WaDr94}\\
\textit{Core (average material properties)}\\
Core radius&\multicolumn{3}{c}{1730\,km}&\citet{Rivo:etal11}\\
Sulfur content&\multicolumn{3}{c}{16\,wt.\%}&\citet{Rivo:etal11}\\
Thermal expansivity\textsuperscript{*,\textdagger}, $\alpha_\mathrm{c}$&\multicolumn{3}{c}{3.5--$4.3\cdot10^{-5}$\,1/K}&\\
Isobaric specific heat, $c_{p\mathrm{c}}$&\multicolumn{3}{c}{750 J/(kg\,K)}&after \citet{StDa08}\\
Thermal conductivity\textsuperscript{*}, $k_\mathrm{c}$&\multicolumn{3}{c}{23.5--25.1\,W/(m\,K)}&\\
\textit{Model impact/crater parameters}&Huygens&Isidis&Utopia&\\
Final crater diameter, $D_\mathrm{f}$&467.25\,km&1352\,km&3380\,km&\citet{JHRobe:etal09,Frey08}\\
Age&3.98\,Gy&3.81--3.96\,Gy&3.8--4.111\,Gy&\citet{Werner08}\\
Impactor diameter\textsuperscript{*}, $D_\mathrm{imp}$&70.6\,km&243.5\,km&699.2\,km&\\
Depth of isobaric core\textsuperscript{*}, $z_\mathrm{ic}$&21.5\,km&74.1\,km&212.9\,km&\\
Radius of isobaric core\textsuperscript{*}, $r_\mathrm{ic}$&19.5\,km&67.1\,km&192.6\,km&\\
\bottomrule
\multicolumn{2}{l}{\textsuperscript{*}Calculated during run. \textsuperscript{\textdagger}Varies with time.}
\end{tabular*}
\end{table}
The density $\varrho$ and other thermoelastic properties of the martian crust and mantle are determined from thermodynamic data for the mineral end-members, which are assumed to form the constituent minerals of the rock as ideal solid solutions. The thermoelastic bulk rock properties are determined in three steps: first, the end-member properties at the desired pressure and temperature are calculated; second, the properties of the individual mineral phases are calculated from them assuming ideal solid solution; third, the bulk rock properties are computed as those of a mixture of the individual phases. In particular, the bulk sound speed $v_\mathrm{B}$ is calculated as the Voigt--Reuss--Hill average of the mineral assemblage \citep{Mavk:etal98} from the densities and the adiabatic bulk moduli $K_S$ of the mineral phases, because our database of thermoelastic properties does not include the shear modulus information necessary to calculate tighter bounds, and theoretical estimates of the shear modulus did not generally achieve an accuracy that would justify their use. The calculation is described in more detail in the Supplementary Material of \citet{RuBr17c}.\par
The electrical conductivity $\sigma$ is determined in two stages from available experimental data on various minerals that are usually not end-members. The conductivity of individual minerals results from the diffusive movement of charge carriers, i.e., of ions (protons or other species) or small polarons (electron--hole pairs) that move through the lattice or along grain boundaries. It can therefore be described by a generalized Arrhenius-type expression of the form
\begin{equation}
\sigma=\sigma_{0i} C_i^{q_i} \exp\left(-\frac{E_i-\alpha_iC_i^{\frac{1}{3}}+p(V_i-\beta_iC_i)}{RT}\right),\label{eq:sig-genArrh}
\end{equation}
where $p$ and $T$ are pressure and temperature, $C_i$ is the concentration of the $i$th charge carrier, $E_i$ and $V_i$ are the activation energy and volume, $\sigma_{0i}$, $q_i$, $\alpha_i$, and $\beta_i$ are material constants, and $R$ is the molar gas constant; this form unifies various extensions of the classic Arrhenius law that are found in the literature in order to include compositional effects. There are in general several different types of charge carriers in a mineral phase, whose contributions to its total conductivity are additive. Of particular interest are the small polarons that arise from surplus charges introduced by Fe\textsuperscript{3+} and cause a dependence on iron content, and the protons that are formed by the dissociation of trace water and strongly enhance the conductivity especially at temperatures that would be expected in the lithosphere. Our models do not indicate the presence of significant amounts of melt on a large scale in the present-day martian interior, especially not beneath large old impact basins. As we have no reliable constraints on the abundance of ice or water in the near-surface regions either, we limit ourselves to the conduction in nominally anhydrous minerals. Hence we first calculate the conductivities of the individual mineral phases using the parameters compiled in Table~\ref{tab:condpar} and taking into account variations of their iron and water contents due to past melting processes and the partitioning of water between the mineral phases according to empirical partition coefficients, as far as such variations are experimentally constrained \citep[e.g.,][]{Auba:etal08,DRBeRo92,KGran:etal07a,Haur:etal06,MMHirs:etal09,Inou:etal10,Nove:etal14a,OLear:etal10,Tenn:etal09}. In the second step, we compute the bulk conductivity of the rock as the self-consistent effective medium conductivity $\sigma_\mathrm{EM}$ given implicitly by
\begin{equation}
0=\sum\limits_{i=1}^N x_i\frac{\sigma_i-\sigma_\mathrm{EM}}{\sigma_i+2\sigma_\mathrm{EM}}\label{eq:sigEM}
\end{equation}
for a composite of minerals with total conductivities $\sigma_i$ and volume fractions $x_i$ \citep[cf.][]{Bruggeman35,Landauer52,Berryman95}. The porosity of near-surface rock decays with depth due to self-compaction under overburden pressure, which is implemented as a simple empirical exponential function and also correspondingly accounted for in the calculation of $\varrho$, $v_\mathrm{B}$, and $\sigma$.

\section{Results}
\subsection{Dynamical evolution and global features of observables}
All models start from an initial state derived from an adiabatic areotherm modified by melting that produced a primordial crust and an initial random depletion of the entire mantle. The models go through an early stage of very vigorous convection characterized by quite irregular and rapidly changing flow patterns that lasts several hundreds of millions of years before entering a calmer and more orderly regime in which about half a dozen large and approximately equally spaced mantle plumes emerges from the core--mantle boundary (CMB). Melting and crust formation, which initially occurred globally throughout the outermost parts of the mantle, continue on an ever decreasing scale for a few hundreds of millions of years before ceasing for good, with the exception of a few very minor and local eruptions. As a consequence of melting, the mantle becomes depleted in iron oxide and, much more strongly, in the radioactive elements and in water, which are very incompatible and thus become concentrated in the crust that is formed by the extracted melt. 10\% of the water in the melt extracted from the mantle is set to be lost permanently to the atmosphere, assuming that volcanic activity on Mars is predominantly intrusive \citep[e.g.,][]{BlMa16}. In the present-day mantle, the former regular melting region is preserved as a depleted shell in the outermost part of the mantle that has become immersed into the outer thermal boundary layer.\par
\begin{figure*}
\centering
\includegraphics[width=\textwidth]{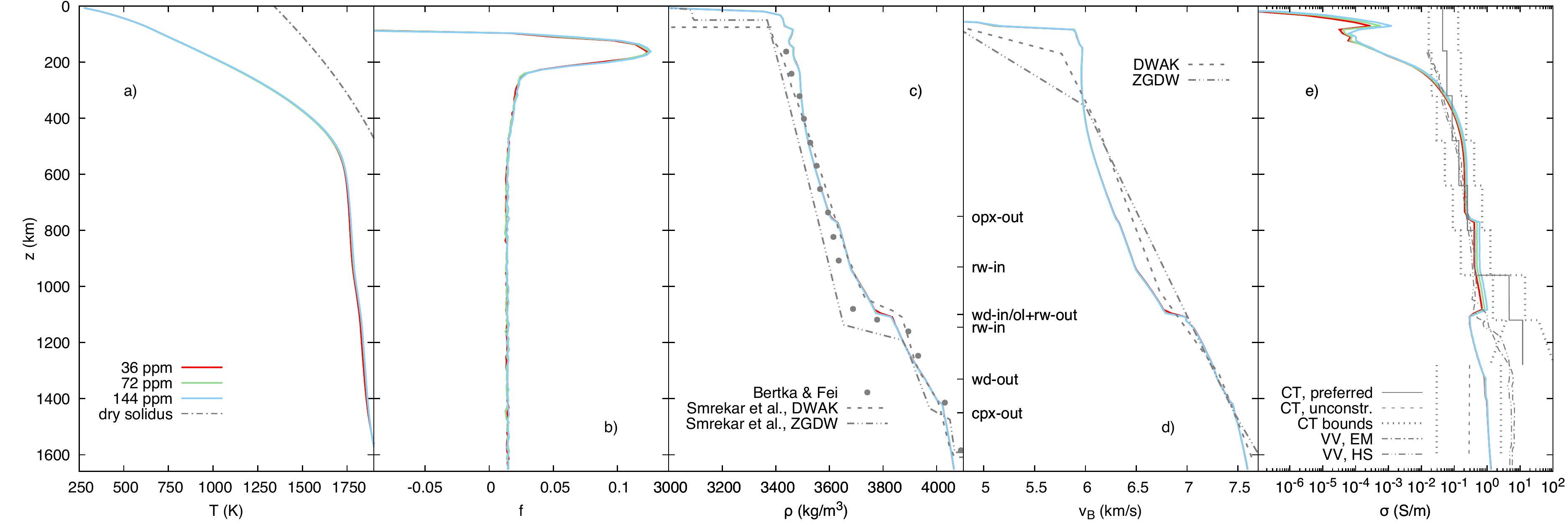}
\caption{Synthetic present-day depth profiles of temperature (a), depletion (b), density (c), bulk sound speed (d), and electrical conductivity (e) of the crust and mantle of the impact-free models with different initial water contents (solid colored curves). The temperature plot also includes the solidus of dry martian peridotite \citep[App.~A]{RuBr17c} (dash-dotted gray curve); at the actual low water contents in the model, the solidus depression would hardly be visible in this plot. Panel (b) also indicates the approximate depth intervals of the crust, the (thermal) lithosphere, and the interval in which regular melting occurred in the ancient past; the approximate position of the crust--mantle boundary is drawn as a dotted line. The gray data points in the density plot (c) are experimentally determined values for a similar composition as assumed by us, taken from \citet{BeFe98} and rescaled to the pressure--depth relation of our models; the gray curves in (c) and (d) are two of the 1D profiles for a composition based on \citet{WaDr94} defined as Mars reference models by \citet{Smre:etal19}. The gray curves in the conductivity plot (e) are preferred inversion results and the bounds on the conductivity from Mars Global Surveyor magnetometer data from \citet{CiTa14} (labeled CT) and the theoretical mineral-physics based models by \citet{VeVa16} (labeled VV) using the effective-medium value as in our model (EM) or the mean of the Hashin--Shtrikman bounds (HS) \citep{HaSh62b}; the data from \citet{CiTa14} are insensitive toward structure at depths of more than 1300\,km. The depths of several phase changes have been marked in (d) to facilitate the correlation with discontinuities in the profiles. The full set of whole-mantle depth profiles for models with and without impacts is available as part of the Supporting Material, Figs.~S6 and S8.}
\label{fig:zref}
\end{figure*}
Fig.~\ref{fig:zref} shows the calculated horizontally averaged present-day depth profiles of temperature, depletion due to melting, density, bulk sound speed, and electrical conductivity, respectively, for the three impact-free models. For comparison, independent observational and modeling data are also shown. Density data for a compositional model based on \citet{WaDr94} (Fig.~\ref{fig:zref}c) have been produced from experiments \citep{BeFe97,BeFe98} or, along with seismic velocity predictions (Fig.~\ref{fig:zref}d), theoretically \citep[compiled by][]{Smre:etal19}. The electrical conductivity (Fig.~\ref{fig:zref}e) has been derived from orbiter magnetometer measurements by \citet{CiTa14} and calculated theoretically for a Wänke--Dreibus model and various laboratory measurements by \citet{VeVa16}. In spite of the different initial bulk water compositions of the models, the profiles are all very similar, because most of the water is partitioned into the crust or lost by outgassing as a consequence of the chosen initial conditions and the melting processes in the uppermost mantle early in the evolution, leaving behind a mantle with at most a few tens of ppm of water even in the most water-rich model after 4.4\,Gy \citep[cf.][]{RuBr17c}. The temperature profiles (Fig.~\ref{fig:zref}a) show that the average temperatures are far below the mantle solidus, which makes it safe for us not to consider the influence of melt on the observables. The depletion ($f$) profiles (Fig.~\ref{fig:zref}b) show that the strongly depleted layer formed by early regular global melting beneath the young lithosphere is now firmly integrated into the lithosphere; the steep shift to negative $f$ marks the boundary between crust and mantle.\par
\begin{figure*}
\includegraphics[width=0.8\textwidth]{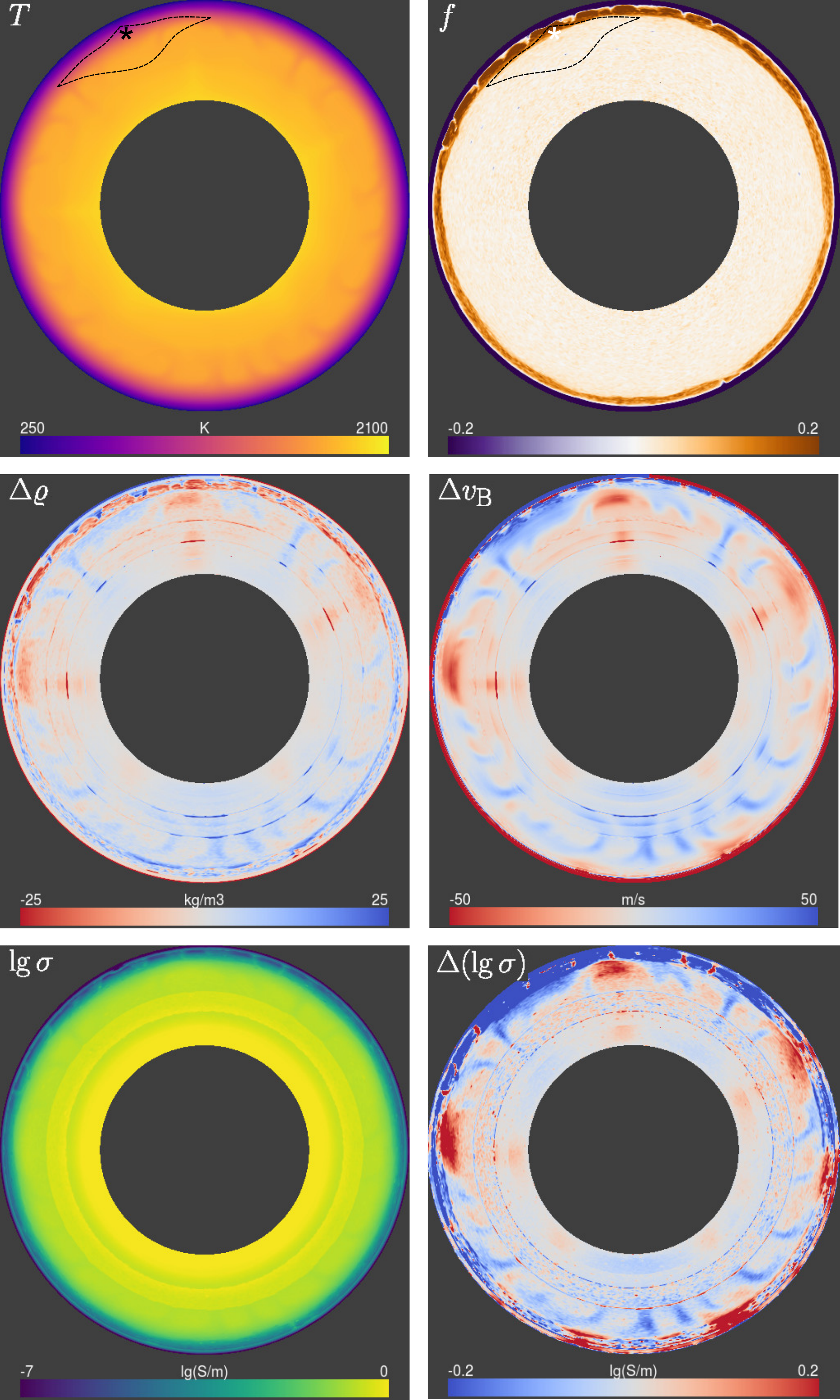}
\caption{From left to right, top to bottom: Present-day temperature $T$, composition/depletion $f$, density anomaly $\Delta\varrho$, bulk sound speed anomaly $\Delta v_\mathrm{B}$, and the decadic logarithm of the electrical conductivity $\sigma$ and the corresponding anomaly, for the Utopia-size impact in a model with 36\,ppm initial bulk water. The approximate position of the center of the isobaric core and of the outline of the original chemical anomaly are marked with an asterisk and a dashed contour in the $T$ and $f$ panels. Except for $T$, the colorbars have been clipped for clarity. Cross-sections of the $T$ and $f$ field at the time of impact and of all fields at the final state are provided for all models in the Supplementary Material, Figs.~S1--S5.}\label{fig:snap}
\end{figure*}

Similar step-like changes can be seen in the profiles of $\varrho$ and $v_\mathrm{B}$; the variations within the crust are mostly due to the closure of pore space with depth and the steep temperature gradient. Smaller changes of $\varrho(z)$ and $v_\mathrm{B}(z)$ can be seen at different depths throughout the mantle and are due to the transformations of olivine to its higher-pressure polymorphs and of the pyroxenes and garnet to majorite; the approximate positions of the transformations are also visible as strong but very thin anomalies at the corresponding depths in Fig.~\ref{fig:snap}, because the Clapeyron slopes of the transitions cause variations in their depth according to the local temperature. The olivine transformations have a more complex structure than in the Earth's transition zone because of an additional appearance of the $\gamma$ phase in the more iron-rich martian mantle, but they are also broader due to the higher Fe content and the shallower lithostatic gradient \citep[cf.][]{KaIt89,Kats:etal04a,Rued:etal13a} and will therefore not cause reflections that are as sharp as in the Earth.\par
The only instance where a clear difference between the models with different water contents can be observed is the lower crust and the uppermost mantle of the electrical conductivity profiles: especially in the lower crust, $\sigma(z)$ differs by up to a factor of 10 as a consequence of the different amounts of water that could be extracted from the mantle source early in the planet's history.

\subsection{Features related to individual impacts}
In models with an impact, the thermal anomaly triggers a pulse of strong, localized melting, which reinvigorates mantle convection on a local or regional scale, depending on the size of the impact and the depth of penetration into the sublithospheric mantle. In the models from \citet{RuBr17c}, in which the impact melt is extracted and forms some sort of crust within the crater, the excessive production of melt also produces a strong compositional anomaly with extreme depletion in incompatible components in the mantle. The thermal and compositional anomaly rises and spreads beneath the lithospheric lid. While the thermal pulse diffuses away with time, the compositional anomaly may persist if it is embedded quickly enough in the growing thermal boundary layer, which is relatively stable due to its high viscosity; this occurs more easily in models with a low water content, because they have generally higher viscosities and convect less vigorously. The final state of this process is shown in Fig.~\ref{fig:snap} (top row) and in Figs.~S2--S5; in the compositional field $f$, where depleted material is visible as regions with a brown hue, the remnants of the impact appear as a dark brown area around the former impact site in the upper part of the cross-section that merges gradually into the undisturbed depleted layer caused by regular melting.\par
In the additional new models of this study, in which impact melt extraction is suppressed, the immediate in situ recrystallization of the impact melt precludes the formation of a similarly strong compositional anomaly, but nonetheless the mantle has been heated up to its solidus in the shocked region. The resulting substantial thermal anomaly still triggers a strong local convective pulse in the aftermath of the impact that undergoes decompression melting to some extent \citep[e.g.,][]{Edwa:etal14} and thus results in the formation of post-impact crust and a compositional anomaly. The latter is weaker than in the cases with extraction and does not spread as far; in terms of global dynamic measures that quantify the vigor of convection such as the root-mean-square velocity of the convecting mantle, the spike-like signature of the impact in models with retained impact melt lies between those with extracted impact melt and models in which all compositional effects were ignored \citep{RuBr17c}. In contrast to the models with extraction, whose final state features an anomaly characterized by extreme depletion, the mantle beneath the basin in models with impact melt retention tends to show no or only slightly stronger depletion than the reference toward the fringes and less depleted material toward the center, which has been drawn into that region by the upwelling thermal anomaly (red and orange curves in Fig.~\ref{fig:zoomHU}b,g; also see the $f$ plots in Fig.~S9 of the supplementary material). The final anomalies in all models are therefore generally broad, flat structures in the uppermost part of the mantle. Significant differences between the models occur only in the uppermost $\sim$500\,km, and so we focus on this depth interval (Fig.~\ref{fig:zoomHU}). In both model variants, we make the assumption that the ejecta that end up within the final crater and the crustal material between the rims of the transient and the final crater form a compositionally homogeneous post-impact crust of constant thickness covering the entire area of the final crater that is emplaced immediately after the impact (computationally speaking, in the same timestep as the impact itself, i.e., within a time span of $\sim$6000 to 18\,000 years). A more refined representation is not possible due to the lack of a sufficiently simple mechanical model and the limited resolution, nor is it necessary, because small-scale features within the crust are not expected to influence the larger-scale effects on the underlying mantle significantly. As a result, we find no major differences in the post-impact crustal thickness in the basins, as far as can be resolved by the vertical grid point spacing of $\sim$13\,km.\par
\begin{figure*}
\centering
\includegraphics[width=\textwidth]{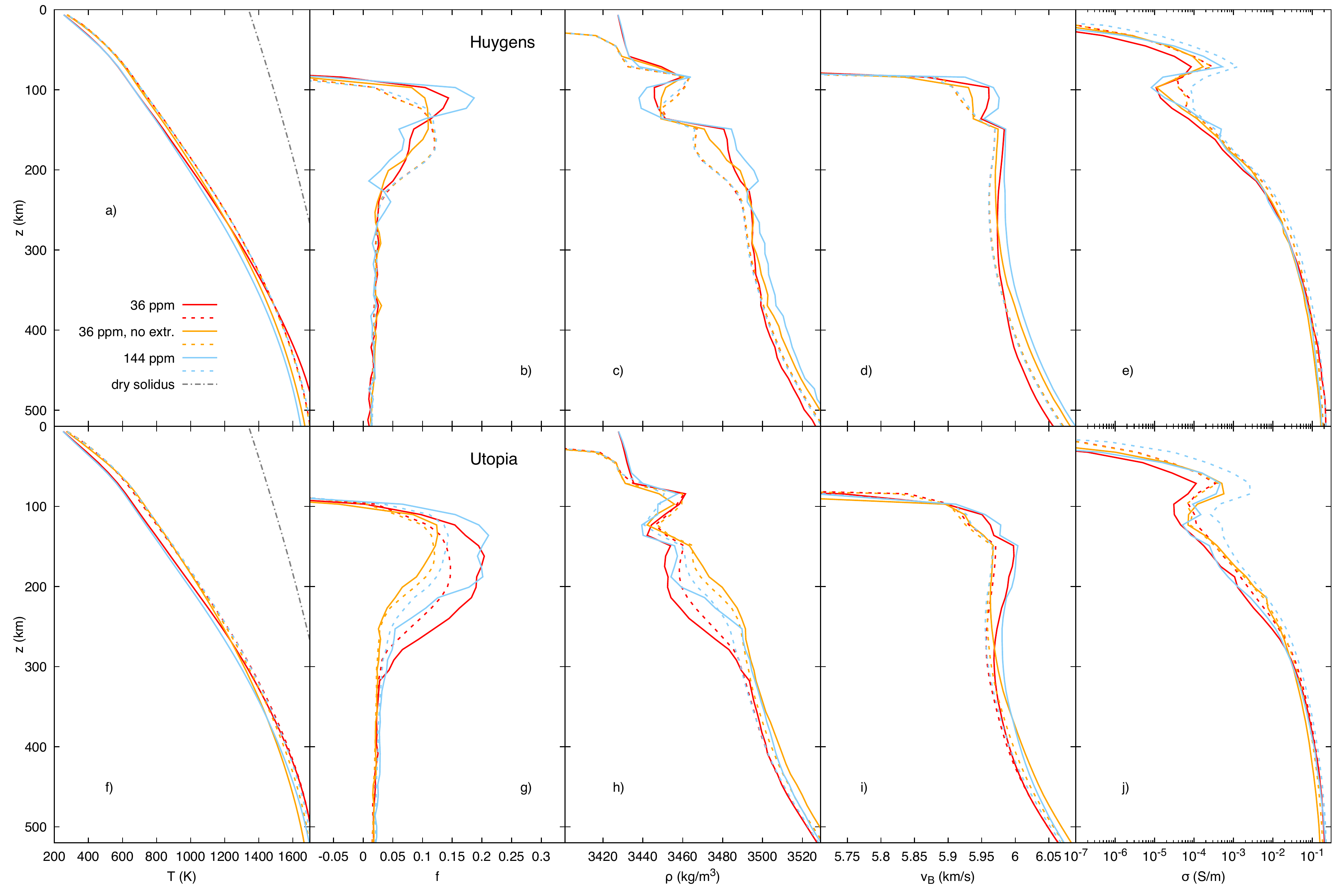}
\caption{Depth profiles of temperature (a, f), depletion (b, g), density (c, h), bulk sound speed (d, i), and electrical conductivity (e, j) in the upper 500\,km of the models with Huygens-size (upper row) and Utopia-size (lower row) impacts. The solid lines are profiles of horizontal averages of the mantle volumes beneath the impact basin, the dashed lines are global horizontal averages. Different colors indicate the different initial water contents of the models (in ppm) and whether impact melt is extracted or not. The full set of profiles, including profiles at the basin centers and the data for models with 72\,ppm initial water, is available as part of the Supporting Material, Figs.~S7 and S9.}
\label{fig:zoomHU}
\end{figure*}
The Huygens impact is the smallest of the three events and barely reaches into the mantle; it is probably a good example for the numerous minor impact basins. It causes a short-lived local disturbance in the uppermost mantle but does not extend the range of melt production to depths greater than those affected by regular melting. Dynamically, its principal effect is to trigger a small upwelling in the lithosphere and in the regular melting zone beneath it. As it spreads beneath the lithosphere and disturbs its structure, it pushes some of the depleted mantle of the melting zone away and sucks in more fertile material from below, but at this time the temperatures are already too low to let it melt. As a result, the mantle anomaly is strongly depleted at the top but shows a decrease in $f$ toward background or even lower values further down, i.e., a blob of unusually fertile mantle sits in its lower part beneath the impact site and becomes frozen in, because the thermal anomaly from the impact dissipates quickly; this blob tends to be better preserved in models with lower water content (cf. Fig.~S2). It has a higher density, mostly as a consequence of its higher iron content. In the upper part, the densities and electrical conductivities are lower but seismic velocities tend to be higher, although the spatial correlation is less sharp, and the deviation affects a larger region. The crust is marked by a pronounced high-density, high-velocity, low-conductivity anomaly at depths shallower than $\sim$100\,km. The effects on density and velocity can be explained by porosity deletion by impact and post-impact melt, whereas the effect on the conductivity seems to be dominated by its temperature dependence: the crust is abnormally cool, mostly because the erupted impact melt is more depleted in radioactive heat sources, such that the crust formed from it has a lower internal heating rate.\par
While the character of the crustal anomaly is a constant of all models with impacts and impact melt extraction, the character of the mantle anomaly is different in the Isidis-sized and the Utopia-sized events, which are substantially larger and clearly affect the sublithospheric mantle. The shock-heated region is large enough to reach beyond the lower margin of the regular melting zone and thus produces an extra volume of depleted mantle that is drawn in by the impact-generated upwelling. As this depleted material has a lower density due to its lower iron content, the corresponding mantle anomaly has a density deficit, contrary to the lower part of the Huygens-sized anomaly. The seismic and conductivity anomalies are somewhat ambiguous with both positive and negative regions, but the positive parts dominate again in the seismic velocities and the negative parts in the conductivities. The anomalies are usually strongest beneath the point of impact and become weaker with distance from that axis, but by and large, depth profiles of the anomaly horizontally averaged over the angular extent of the corresponding impact basin share many features of the corresponding central profiles and tend to appear as somewhat muted and smoothed versions of them (cf. Figs.~S6--S9).

\subsection{General impact-related features}
\subsubsection{Electrical conductivity}
The electrical conductivity is sensitive to the changes in water and iron contents induced by melting and displays contrasts of more than an order of magnitude between anomaly and reference in some cases (Fig.~\ref{fig:zoomHU}, rightmost column). While all profiles are fairly similar in the deeper mantle, because even in the most water-rich models the mantle has retained only a few tens of ppm water at most, the impact-generated anomaly is marked by a stronger conductivity minimum than parts of the mantle that have only experienced regular melting. Given the generally low water content of the mantle in all models, these contrasts are mostly due to stronger depletion in iron and the smaller fraction of surviving pyroxene in the impact anomaly, where much higher melting degrees had been reached. The conductivity also reveals more physical structure within the crust than the other methods considered: while they are mostly sensitive to the porosity, $\sigma(z)$ bears the mark of further influences and is therefore more variable between models. Compared to the mantle, the crust is characterized by a marked increase of the conductivity across the compositional boundary between the two, which is due to the mineralogy and the higher iron and water contents in the crust ($\mathrm{Mg\#}\approx0.47$ and H\textsubscript{2}O on the order of several hundred to a few thousand ppm). In particular the crustal water concentrations, which vary between models according to the initial water contents, result in visible conductivity differences between the models; these systematic differences with water content are most clearly expressed in the global average profiles, whereas the profiles at the basin center may be influenced by local conditions. On the other hand, the strong temperature dependence of $\sigma$ reflects the slightly lower temperatures in the basin crust and comes to dominate at shallower depth in combination with the increasing porosity.
\subsubsection{Seismic velocities and density}
In contrast to the electrical anomalies, the bulk sound speed and the density anomalies in the mantle are fairly small. The $v_\mathrm{B}$ anomalies are mostly positive and reach the largest values of up to 100\,m/s in the uppermost mantle but drop to generally less than 50\,m/s beyond 150\,km depth; beneath the Huygens basin, the anomalies are generally a bit weaker due to the lesser size of the impact. The crust in the basin is always seismically faster in the models with impact melt extraction because of the closure of pore space (Fig.~\ref{fig:zoomHU}). In terms of seismic travel times, this means that seismic arrivals are earlier than in the global average, but if impact melt is not extracted, the travel time differences are insignificant. To get an estimate of the travel time differences, we calculated them for the simple configuration of a vertical ray entering an anomaly consisting of layers of the thickness of the numerical grid shells and spanning the depth range of the uppermost 500\,km of Mars beneath the basin. For a stack of $N$ layers of thickness $\Delta z_i$ with density $\varrho_i$, adiabatic bulk modulus $K_{Si}$, and bulk sound speed
\begin{equation}
v_{\mathrm{B}i}=\sqrt{\frac{K_{Si}}{\varrho_i}}=\sqrt{v_{\mathrm{P}i}^2-\frac{4}{3}v_{\mathrm{S}i}^2},
\end{equation}
the travel time difference for a vertical ray with respect to a reference at the same depth level is
\begin{align}
\Delta t_\mathrm{B}&=\sum\limits_{i=1}^N \Delta z_i\left(\sqrt{\frac{\varrho_i}{K_{Si}}}-\sqrt{\frac{\varrho_{\mathrm{ref},i}}{K_{S\mathrm{ref},i}}}\right)\nonumber\\
&=\sum\limits_{i=1}^N \frac{1}{\sqrt{\frac{1}{t_{\mathrm{P}i}^2}-\frac{4}{3t_{\mathrm{S}i}^2}}}-\frac{1}{\sqrt{\frac{1}{t_{\mathrm{Pref},i}^2}-\frac{4}{3t_{\mathrm{Sref},i}^2}}},\label{eq:dtB}
\end{align}
where $t_{\mathrm{P,S}i}$ are the travel times of P- and S-waves in the $i$th layer; the latter form is more useful for comparison with observed travel times, because the travel times for $v_\mathrm{B}$ cannot be determined directly from seismograms. The estimates for a vertical ray passing through an anomaly with the properties of the laterally averaged mantle beneath an impact basin covering the upper $\sim$500\,km of the model are given in Table~\ref{tab:travtime}. Applying the rule of thumb that $v_\mathrm{P}=\sqrt{3}v_\mathrm{S}$ (Poisson solid) implies that $v_\mathrm{B}=\sqrt{5/9}v_\mathrm{P}=\sqrt{5/3}v_\mathrm{S}$; multiplying the travel time anomalies in the table with $\sqrt{5/9}\approx0.745$ or $\sqrt{5/3}\approx1.291$ therefore gives a more accurate estimate of the travel time anomalies for the P- and S-waves, respectively. For an anomaly with the bulk sound speed profile beneath the point of impact or with the laterally averaged $v_\mathrm{B}(z)$ beneath the basin, we find that such seismic waves would arrive several seconds earlier within the basin. More than 95\% of the speed-up occurs in the crust, and the effect is the stronger the smaller the basin is.\par
\begin{table*}
\centering
\caption{Estimates of travel time anomalies for bulk sound speed (Eq.~\ref{eq:dtB}) in seconds.}\label{tab:travtime}
\begin{tabular}{lcccc}\hline
Model&\multicolumn{2}{c}{Mantle}&\multicolumn{2}{c}{Total}\\
&basin center&average&basin center&average\\\hline
Huygens, 36\,ppm&$-0.12$&$-0.08$&$-10.53$&$-10.62$\\
\hspace*{2em}no melt extraction&$-0.16$&$-0.1$&$-0.49$&$-0.26$\\
Huygens, 72\,ppm&$-0.15$&$-0.14$&$-11.04$&$-10.97$\\
Huygens, 144\,ppm&$-0.33$&$-0.28$&$-10.89$&$-10.86$\\
Isidis, 36\,ppm&$0.01$&$0.04$&$-7.73$&$-7.88$\\
\hspace*{2em}no melt extraction&$-0.12$&$-0.06$&$-0.29$&$-0.06$\\
Isidis, 72\,ppm&$-0.48$&$-0.34$&$-8.56$&$-8.28$\\
Isidis, 144\,ppm&$-0.35$&$-0.28$&$-8.11$&$-8.01$\\
Utopia, 36\,ppm&$-0.31$&$-0.12$&$-4.94$&$-5.33$\\
\hspace*{2em}no melt extraction&$-0.12$&$-0.1$&$-0.2$&$-0.01$\\
Utopia, 72\,ppm&$0.07$&$-0.22$&$-4.89$&$-5.50$\\
Utopia, 144\,ppm&$-0.46$&$-0.25$&$-5.76$&$-5.51$\\
\hline
\end{tabular}
\end{table*}
As pointed out in our earlier paper \citep{RuBr17c}, the density anomalies in the mantle do not exceed a few tens of kg/m\textsuperscript{3} and are mainly a consequence of the depletion in iron and the disappearance of dense minerals, especially garnet. The density contrast in the crust can be up to an order of magnitude larger, mostly due to the deletion of pore space within the basin. Especially the shallower parts of the mantle anomalies are mostly negative for these compositional reasons, whereas they are positive or variable at depths greater than 200\,km. This is the case in the Huygens models, where the impact effects were mostly confined to lithospheric depths anyway, and also in the more water-rich models of the larger impacts, but the variations are so small that it would be difficult to distinguish impact-related effects from the normal local variability in a convecting mantle.

\section{Discussion}
Our survey indicates that the geophysical characterization of impact-generated anomalies may in principle be possible, but various limitations and ambiguities have to be considered.
\subsection{Electrical conductivity}
Electrical conductivity distinguishes itself from the other variables discussed here by its stronger sensitivity to temperature, mineralogy, iron content, and water concentration. Our calculations indicate that local structures with unusually low conductivity may be detectable and that one may even be able to discern to some extent between models in spite of the low water contents. An analysis of the sensitivity to these factors (cf. App.~\ref{app:sens}) shows that the sensitivity to temperature is significant everywhere and that the temperature governs the overall conductivity structure in combination with the basic mineralogical composition. At several per cent per kelvin or more, the sensitivity to $T$ in terms of the logarithmic temperature derivative of $\sigma$, which is itself temperature-dependent, is strongest at low temperatures but still significant at least in parts of the mantle (cf. Fig.~\ref{fig:sens-glob}). The influence of mineralogy is most obvious in the contrast between crust and mantle and in some shifts in the profile that coincide with phase transitions. In terms of relative change, $\sigma$ is also very sensitive to changes in water concentration, especially at low temperatures and/or concentrations, but in order to detect substantial variations in water content via conductivity, reference concentrations of at least several tens of ppm are necessary (cf. Figs.~\ref{fig:zref}, \ref{fig:sens-glob}, S11, and S12).\par
In order to assess our models, we compare the global means with the 1D profile from \citet{CiTa14}, which was derived by inversion of MGS magnetometer measurements using a model with nine 180\,km-thick layers (Fig.~\ref{fig:zref}, rightmost column). We find quite good agreement between our models, their MGS data inversion profile, and the conductivity model constructed from a purely mineralogical framework by \citet{VeVa16} between $\sim$300 and $\sim$1000\,km depth corresponding to the olivine stability field beneath the regular melting zone, in spite of the different conductivity databases used and different assumptions about the distribution of water between the mineral phases. Our models are also within the bounds of the MGS model in the lowermost mantle ($z\gtrsim 1300$\,km), where the observational data are insensitive to structure, but it predicts lower $\sigma$ than the inversion of the MGS data and \citet{VeVa16} in the olivine+ringwoodite and the wadsleyite shells in between; older synthetic profiles by \citet{MoMe00} generally yield too high conductivities. The sensitivity tests suggest that water contents in that depth range would have to be more than a hundred times higher than in our models to match the conductivity inferred from MGS data.\par
The synthetic models all show a steep decline in $\sigma$ above $\sim$300\,km that is not seen in the MGS inversion, which has quite high conductivities up to the surface. Similar conductivities (0.1--0.05\,S/m) are also found for periods with penetration depths of 100--200\,km in InSight magnetometer data \citep{AMitt:etal20a}. The MGS model lumps together the crust and much of the regular melting zone into one single layer and therefore cannot resolve the internal structure of this depth range; it is hence expected that it does not display such a steep decline. However, the generally high conductivity in that uppermost layer indicates that the MGS data include contributions not considered in our conductivity calculation. One likely possibility is the existence of impure water ice and/or brines, which were detected \citep[e.g.,][]{Piqu:etal19} in the pore space near the surface and may contribute substantially to the bulk conductivity; the content of water (in whatever form) near the surface has been estimated as ranging from $\sim$3\% at low latitudes to $\sim$13\% in north polar regions \citep[e.g.,][]{Audo:etal14}, and substantial amounts of water at depths of a few kilometers are also corroborated by the high-frequency part of the InSight magnetometer measurements \citep{YYu:etal20}. While one might intuitively expect that another contribution comes from hydrous minerals, which have been detected by spectrometers on Mars rovers and orbiters \citep[e.g.,][]{Cart:etal13} as well as meteorites \citep[e.g.,][]{McCu:etal16a}, the relatively scarce available experimental data for such minerals as amphibole and talc do not support this. Although the presence of amphibole could increase the conductivity of the bulk rock to some extent, the strongest increase in conductivity occurs upon its dehydration \citep[e.g.,][]{Fuji:etal11,DWang:etal12}, and so its effect would be most noticeable at temperatures above $\sim$800\,K, but this does not seem to occur in the modern martian crust (cf. Fig.~\ref{fig:zref}). The conductivities of talc or rock like serpentine are not conspicuously high \citep[e.g.,][]{XGuo:etal11}. Therefore, we conclude that the shallow high-conductivity structure, if it is confirmed by independent measurements, is essentially due to brines and impure ice, and no significant error is introduced in our calculation by omitting them.\par
An impact site is expected to stick out as a region with particularly low conductivities in the uppermost mantle and also a reduced conductivity in its crust, but even its conspicuousness does not ensure that it can be distinguished with certainty from the site of a former mantle plume with no relation to an impact; the final verdict will usually require the combination of several geophysical observations and clues from surface geology. The low crustal conductivity within the crater is mostly a consequence of lower temperature and reduced water content in our models and is rather partly compensated by post-impact porosity deletion in models with impact melt extraction; such porosity-related effects would be limited to the uppermost layer of our numerical grid, because the vertical grid point spacing is $\sim 13$\,km, and overburden pressure reduces the porosity to about 0.1\% at 40\,km depth. However, if we had included ice or brines, which have a relatively high $\sigma$ compared with silicates at the same temperature, in near-surface pore space in our models, then the removal of pores by impact melts reduces the pore space available for these materials, and the calculated net effect would rather reinforce the reduction of the conductivity relative to the reference in the topmost few kilometers. It is generally at this very shallow crustal level where their effect would be significant, i.e., at depths not resolved in our model, and it is therefore of little relevance within the scope of this study, as the large-scale structures we model would be targeted by more deeply penetrating very-long period parts of the electromagnetic spectrum that are less sensitive to fine-scale shallow structure. Nonetheless, an inversion model of electromagnetic data for the deep interior would have to take the effect of near-surface conductivity structure into account but should construct it independently, e.g., on the basis of SHARAD or MARSIS data \citep[e.g.,][]{Oros:etal15,Seu:etal04} or similar but more deeply penetrating measurements; the maximum penetration depths of SHARAD and MARSIS are 1 and 5\,km, respectively \citep{Seu:etal04}. This would constrain the properties of the upper crust, in particular the thickness and conductivity of sedimentary layers formed by later post-impact processes that are not covered at all in our model. By and large, this prediction of conductivity anomalies on large scales and the depth range of the whole mantle and their detectability should hold up in spite of the very considerable discrepancies between mineral conductivities measured in different laboratories, which were discussed in detail by \citet{AGJones14a,AGJones16} especially with regard to olivine.\par
The conductivity profile constructed by inversion of MGS data will be representative for the laterally averaged electrical conductivity of the martian interior and can thus serve as a background reference comparable to the dashed profiles in Figs.~\ref{fig:zref} and \ref{fig:zoomHU}, but it will probably not be possible to extract regional features from such data. In order to detect an impact-generated anomaly, it would be necessary to put a station on the ground, ideally at the center of the expected anomaly, which would have to be able to collect contiguous data series over weeks or months in order to cover the very long, deeply penetrating periods in the EM spectrum that can probe the entire depth range of the mantle. As even the measurements from a single station can facilitate the construction of a local one-dimensional $\sigma(z)$ profile, comparison of such a profile with the MGS-derived global average should reveal existing anomalies. The case for an impact as the cause of it would of course be strengthened if additional stations on the ground were available. At least one of them should be located in an area remote from any large impact basin, volcanic province or similar prominent features, i.e., in a region that is as boring as possible. That station could help to validate the MGS-derived reference, because it should not display strong deviations from it that are due to geodynamical processes other than impacts.

\subsection{Seismic velocities and density}
One of the benefits to be reaped from seismic data is structural information, which is one of the reasons it is part of the ongoing InSight mission. Indeed, the calculated $v_\mathrm{B}(z)$ profiles show major discontinuities such as the crust--mantle boundary and those at the mid-mantle phase transitions; the predicted velocity jump across the CMB would be approximately 4.43\,km/s. Some of the mid-mantle transitions are probably rather smooth, because the transitions span a larger pressure interval than in the terrestrial mantle because of the wider two-phase loops of olivine at Mg\#=0.75 and because of the shallower lithostatic pressure gradient. Although Mg\# for Mars seems to be fairly well constrained by petrological studies, high-quality seismic data in which the depths and widths of the phase transitions are identifiable should confirm this value via the pressure--depth relation. At the same time, this information would also put constraints on the temperature of the mantle and thus help to reduce the trade-off between important factors that affect the observed electrical conductivity. This would be especially helpful with regard to the temperature sensitivity test of the density, which showed that a slightly higher temperature in the mid-mantle would reproduce the experimental densities from \citet{BeFe98} even better between 750 and 1150\,km depth than the profile from Fig.~\ref{fig:zref} (cf. Fig.~\ref{fig:sens-glob}), whereas the agreement of the nominal models with the inversion model DWAK from the compilation of \citet{Smre:etal19} is quite good in the density profiles. The bulk sound speed profiles agree less well, but the comparison confirms that model DWAK is more in line with our models than model ZGDW, although all three are based on the Dreibus--Wänke compositional model; a more definitive judgment will have to wait until results from InSight become available. There is a trade-off between the thermal and the compositional contribution to a density anomaly that cannot be resolved with gravity information alone; as an example, in our mineral physics model, a density reduction of primitive mantle material by 45\,kg/m\textsuperscript{3} can correspond to a temperature increase of some 350\,K or an increase in Mg\# by about 0.05. In principle, the situation would be complicated further by varying water contents, but at the low water concentrations expected especially in the mantle, it is safe to neglect this effect; judging from the change in molar volume by hydration, which is between 5 and 25\% for the relevant minerals \citep[table~2]{KeBo-Ca06} the expected relative density changes are on the order of $10^{-4}$ or less. As with the density, we do not expect seismics to be able to constrain the water content, because at the low concentrations, the expected reduction of the bulk modulus is on the order of a few tens of MPa or less \citep[eq.~3]{Jacobsen06}, well below the uncertainty of bulk modulus data.\par
More subtle variations such as the impact-generated anomaly, however, will be much harder to detect and interpret, because the velocity changes are mostly not very sharp or clearly structured, and at only a few tens of m/s, their magnitude is rather small. We tested the sensitivity of $v_\mathrm{B}$ to variations in $T$ and $f$ (cf. App.~\ref{app:sens-seis}) and found them to have opposite signs. The tests show that in the uppermost mantle, a twofold increase in $f$ as might be typical for an old impact-generated anomaly would appear similar in magnitude as a 10\% reduction of temperature (Fig.~\ref{fig:sens-glob}), but it would have limited depth extent over a large area, which would be less likely (although not impossible) for a purely thermal cold anomaly for dynamical reasons. The fact that $v_\mathrm{B}$ is more sensitive to $T$ variations than the density is of little use in the context of impact-generated anomalies, because the thermal anomaly has long diffused away. Hence, these considerations seem to confirm our conclusion from our previous paper that the impact-related anomalies could at best be detected by a targeted regional seismic network, as far as seismic methods are concerned. On the other hand, one should keep in mind that our choice of $v_\mathrm{B}$ was born out of the necessity to construct a seismic observable from a thermoelastic database without shear moduli. From a practical point of view, S-wave velocities would be a direct result of seismic measurements and do not require good coverage of the target volume by both P-wave and S-wave raypaths, as would be necessary for determining $v_\mathrm{B}$. Generally, our assessment leaves out various other possibilities offered by the analysis of seismic waves, e.g., the investigation of anisotropy or attenuation. Changes in these properties, which are related to variations in mineralogy and water content, could at least in principle further assist in the identification and characterization of impact-generated anomalies.\par
Maybe the analysis of surface waves reaching an off-basin station from different directions would also give hints, because the basin should appear as a large fast, roughly symmetric anomaly and possibly show a different anisotropy pattern than its surroundings. Specifically, surface waves reaching the InSight lander from northerly and westerly directions should be affected by the Utopia and Isidis basins, respectively, whereas steeply incoming body waves are presumably passing through a medium similar to the global mean.

\subsection{Melt extraction}\label{sect:meltextr}
As already remarked in the Introduction, our preferred implementation of impact melt formation and dynamics and the corresponding synthetic observable calculations depend on the assumption of efficient melt segregation from the shock-molten region. The effects we predict for our models with impact melt extraction are therefore the most optimistic scenario, whereas models with some impact melt retention in the mantle, which may be more realistic, show diminished effects. A full treatment of melt dynamics during crater formation would require the combination of a fully dynamical impact model with a two-phase flow algorithm and is beyond the scope of our model. Therefore, we limit ourselves to some considerations of the length and time scales relevant to the segregation of melt in those regions of the mantle surrounding the transient crater that are partially molten by the shock wave. A lower limit on the segregation velocity of melt may be estimated by assuming an isotropic medium with grains of diameter $a$ and melt-filled pores, whose permeability can be approximated by the generalized Kozeny--Carman relation $k_\varphi\approx a^2\varphi^n/b$. Of interest are chiefly regions with porosities $\varphi$ of at least a few per cent, for which the melt can be assumed to form a network of isotropically oriented interconnected films; in this case, the geometrical factor $b$ is 162 and $n=3$ \citep[e.g.,][]{ScMa14}. The melt segregation velocity applicable to Darcy flow is
\begin{equation}
\vec{u}_\mathrm{seg}=\vec{u}_\mathrm{f}-\vec{u}_\mathrm{s}=-\frac{k_\varphi}{\eta_\mathrm{f}\varphi}\nabla p
\end{equation}
\citep{McKenzie84}, where $\vec{u}_\mathrm{s,f}$ are the velocities of the solid and the fluid phase of the partially molten rock, $\eta_\mathrm{f}$ is the viscosity of the magma, and $p$ is the pressure that drives the flow. Under the very simplistic assumption that the only pressure driving the flow comes from the weight of the melt column above a certain depth level $z$ and acts over a lateral distance $d$ from the crater wall, we have then
\begin{equation}
\vec{u}_\mathrm{seg}=-\frac{a^2}{b\eta_\mathrm{f}}\varphi^{n-1}\frac{\varrho_\mathrm{f}gz}{d}.
\end{equation}
With $a=1$\,cm \citep[e.g.,][]{HiKo03}, $\eta_\mathrm{f}$ on the order of 0.1--1\,Pa\,s \citep[e.g.,][]{HNi:etal15}, $\varrho_\mathrm{f}=2750$\,kg/m\textsuperscript{3}, and $d=10$\,km, we find that melt can flow with velocities of at most several millimeters per second into a transient crater on Mars with a depth of up to a few tens of kilometers.\par
However, it is likely that the crater floor is shattered and criss-crossed by large open cracks with widths ranging from centimeters to meters in which melt or magma can be transported much more efficiently. We applied a recent analytical model developed by \citet{HeWi20a} to model the injection of lunar magmas into existing fractures to try to estimate an upper bound for the distances and velocities of melts flowing from the surroundings of the crater walls toward the transient crater. The material properties were changed from the original to values applicable to a basaltic magma with the composition of the martian crust \citep[tab.6.4]{SRTaMcLe09} using the melt viscosity model by \citet{Gior:etal08} along with densities after \citet{RALaCa90} and \citet{KrCa91} and heat capacities from \citet{LeSp15}. The model distinguishes between an initial stage in which the magma flows turbulently in a fracture of constant width and a final stage with laminar flow in a fracture that narrows as the magma flow solidifies from the walls inward; the turbulent stage may not occur in narrow fractures with widths of a few centimeters or less. In the turbulent flow regime, loss of heat from the melt through the fracture walls raises the viscosity of the liquid and triggers crystallization at some point. The latter process leads to a marked and progressive increase in the viscosity of the magma that finally results in the transition to laminar flow. In the calculation of the laminar stage, we consider only the state at the central plane of the fracture, which is the last part to reach a yield strength too high to allow for further flow; this implies that the travel distances for the magma are maximum values and a substantial fraction of the magma will stall much earlier. We use the same rough estimate for the pressure driving the flow and find that in the turbulent flow velocities of tens of meters per second may be reached in decimeter-wide fractures but the distance covered does not exceed a few hundred meters even then; the actual distance depends on the initial temperature of the melt. By contrast, the flow in the laminar stage is on the order of decimeters per second to a few meters per second but can be maintained over distances of kilometers or tens of kilometers in fractures with widths on the order of decimeters \citep[after][]{PiSt92,PiWi94,Roscoe52}. Melts with a dunitic composition that flow through hotter host rock have viscosities that are one or two orders of magnitude lower \citep{Pers:etal18b} and are cooled less efficiently, which allows them to travel even larger distances. In the study of \citet{HeWi20a}, which is aimed at magmatic intrusion under normal conditions from a source at depth, the case is made that most cracks observed in the field, including those in impact structures, do not exceed a few centimeters in width, and thus melt transport would be much less efficient than calculated here. We do not dispute the validity of that reasoning but argue that in the highly dynamic environment of an opening crater, wider fractures may form as transient structures that would not be preserved in the geologic record but would nonetheless allow high-volume flow for short timespans, and the driving pressures encountered in the neighborhood of a slumping crater wall could be higher than our conservative guess. In summary, the estimate suggests that in principle velocities are possible that could drain melt from a mantle zone around the crater of several kilometers into the cavity within a few minutes, which is of the order of magnitude of the time scale of the opening, collapse, and rebound of a large impact basin.\par
In a more realistic scenario, the pressure and shear stresses exerted by the compacting matrix and the collapsing crater walls may help to squeeze the melt out, although it likely depends on the exact structure and temporal evolution of the stress field whether melt in the vicinity of the crater is extracted or trapped. A two-phase flow model of impact--melt dynamics does not yet exist to our knowledge, but \citet{Mans:etal19a,Mans:etal21} have recently developed impact models in which the production and transport of non-segregating melt is tracked. Such models can give useful indications about the fate of melt that is trapped or segregates only slowly on timescales longer than that of crater formation. Of their models, those for a young Mars hit by an impactor with 30 or 300\,km diameter can be compared with our Huygens and Isidis models, respectively. They show that in general the melting degree increases with decreasing depth, causing a major part of the produced melt to be relatively close to the surface and hence likely to extrude; this is partly due to the lower solidus of crustal material compared to that of depleted mantle. By contrast, a certain fraction of the melt in the less pervasively molten regions at greater depths, especially in the mantle, may remain trapped and recrystallize if the prevailing stress field prevents its efficient expulsion. In the bigger impact, which affects a larger part of the already hot and hence more easily fusible mantle, a large volume undergoes extensive melting to intermediate or high degrees, and a substantial part of this melt should reach the surface in the immediate aftermath of the impact. The crucial variable is thus not the absolute size of the impactor or the depth of the opening crater but their size relative to the thickness of the cool crust or lithosphere, which increases as the planet ages. In a Utopia-size impact, this effect would presumably be even stronger.\par
A large volume of extracted melt would compensate the removal of crust by the impact itself and could even result in thickened crust with low porosity, as it does in most of our models. In \citet{RuBr17c}, we calculated along-axis gravity anomalies for a simple model with cylindrical symmetry to estimate the order of magnitude of the crust on the gravity signal and found a positive anomaly on the order of 100--200\,mgal on the surface that is strongly reduced or absent in models without melt extraction. This is consistent with the observed Bouguer anomalies \citep[e.g.,][]{Geno:etal16}, although the amplitude is not large enough to explain the signal entirely. However, our treatment of all erupted melt as ``basaltic'' even at very high melting degrees and the exclusion of processes such as the differentiation of a post-impact magma pond is likely to result in an overestimate of the calculated crustal thickness and the anomalies related to the crust. Field studies of large impact structures suggest that such magma ponds differentiate to some extent \citep[e.g.,][]{IvDe99,Laty:etal19}, and various petrological studies found evidence for differentiation in the martian crust \citep[e.g.,][]{Ruiz:etal06b,Bara:etal14,Sant:etal15,Saut:etal15,Udry:etal18}, although they do not always relate it strictly to impact events. If the denser lower parts of such a large differentiated magma pond have densities between that of normal dense crust and that of depleted mantle, they would be difficult to distinguish from an uplifted mantle body by their gravity signature. The case for a thick, differentiated crystallized melt sheet with partly ultramafic composition in the subsurface of large basins has been made in particular for the South Pole--Aitken basin on the Moon \citep[e.g.,][]{RNaka:etal09,VaHe14}; indeed, the latter authors proposed a top--down progression from a noritic to a dunitic lithology for the differentiated post-impact melt sheet of that basin with a total thickness of 50\,km, i.e., the lowermost segment would consist of a mantle-like material. Another possibility is that high-degree and/or ultramafic melts from the deeper reaches of the shocked region get trapped somewhere at depth; we should emphasize that in reality ``extraction of melt'' only means that melt is removed from its source region but not necessarily that it reaches the uppermost few kilometers or even the surface, even though we have implemented it this way in our models for numerical reasons. Either way, the boundary between crust and mantle in such large impact basins may be blurred and have the character of a more gradual transition than in undisturbed regions. In seismic data, such a structure could manifest itself for instance as a weak or absent reflection from the martian Moho. If the boundary is indeed not sharp, efficient melt extraction would not necessarily imply a large crustal thickness in the conventional sense, nor would a gravity high necessarily imply a thinner crust. Instead, the contradiction with the existence of a thin crust that is inferred for many basins would merely be a consequence of the clear distinction between crust and mantle that is made by the inversion models but that is not appropriate here.

\subsection{Additional considerations about multiple impacts and global structure}
\begin{figure*}
\includegraphics[width=0.8\textwidth]{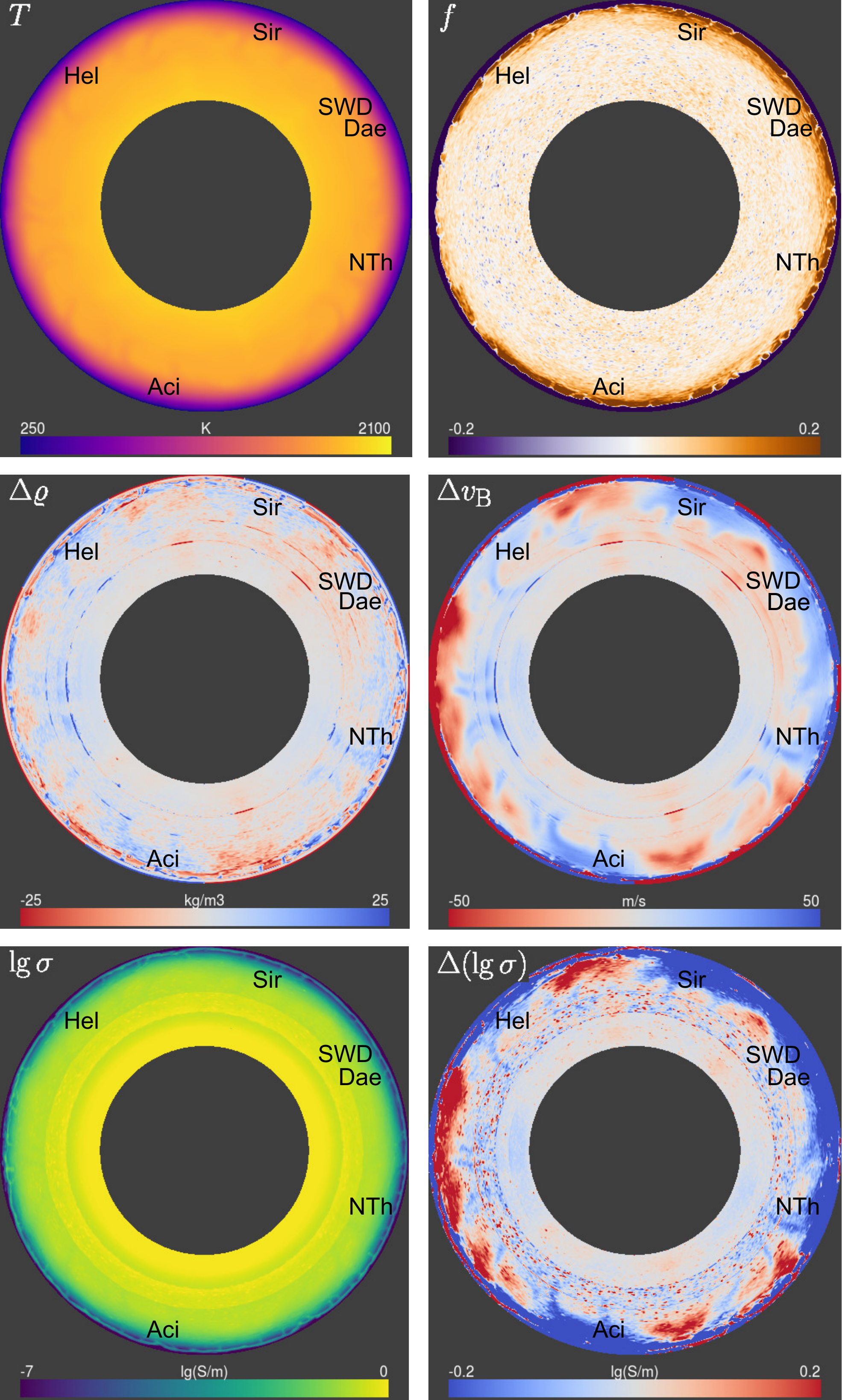}
\caption{From left to right, top to bottom: Present-day temperature $T$, composition/depletion $f$, density anomaly $\Delta\varrho$, bulk sound speed anomaly $\Delta v_\mathrm{B}$, and the decadic logarithm of the electrical conductivity $\sigma$ and the corresponding anomaly, for model GC4 from \citep{RuBr19a}, which has six impacts (Hel = Hellas; Sir = Sirenum; SWD = Southwest of Daedalia; Dae = Daedalia; NTh = North of Tharsis; Aci = Acidalia). Except for $T$, the colorbars have been clipped for clarity.}\label{fig:multi}
\end{figure*}

In the interpretation of observed anomalies it should be kept in mind that planets were not hit by one single basin-forming impact, and models with multiple sufficiently closely spaced impacts show that their signatures may overlap \citep[e.g.,][]{RuBr19a}; indeed, the mere existence of several large impacts will already leave a mark on the global means that serve as references that reduces the signature of individual anomalies. To get an impression of what to expect in such a setting, we calculated anomalies for models with 4, 6, or 8 successive basin-forming impacts of different size from \citet[fig.~7]{RuBr19a}; Fig.~\ref{fig:multi} shows the same variables as Fig.~\ref{fig:snap} for a six-impact case (GC4) from that study (see the Supplementary Material for the present paper for more examples), but one must keep in mind that geometrical effects give excessive weight to local anomalies like impacts in the calculation of mean quantities in two-dimensional models as compared to three-dimensional reality, as discussed in that paper. Those additional examples show how the merging and overlapping of the compositional anomalies of multiple basin-forming impacts extends to the geophysical anomalies: impact-affected regions still stick out quite clearly especially as seismic and electrical conductivity anomalies, but there are no sharp boundaries; rather, overlapping compositional anomalies are visible in those variables as one large negative anomaly with varying depth extent. One must keep in mind, however, that these are two-dimensional models; in three dimensions, an impact anomaly would have to be completely surrounded by neighboring anomalies to have its margins blurred by merging and overlapping, which is much less likely for very large basin-forming impacts.\par
A special problem is the giant impact that resulted in the formation of the Borealis lowlands, because it was a singular event of a magnitude substantially larger than even the Utopia-forming impact and may have had global effects. Our method is not suited to modeling impacts of this scale, but one can attempt an assessment on the basis of dynamical calculations of the impact such as those by \citet{Mari:etal08,Mari:etal11}. Those models do not cover the long-term evolution, but the extraordinary depth extent of the strongly shocked region and the size and melting degree of the partially molten region in the impacts with the highest energies and steepest angles suggest that a large fraction of the martian mantle would have been at least partially molten, partly also due to additional melting at the antipodes of the impact point. The resulting thermal and compositional anomaly would in principle have behaved similarly to the smaller ones in our models, i.e., it would have ascended and spread out at the top of the mantle beyond the region visible at the surface as the dichotomy boundary. However, the models from those studies that give the best match with the observed dichotomy point to a moderately oblique impact that would have produced much lesser effects. Therefore we conclude that the anomaly from a possible dichotomy-forming impact, which would have happened very early in Mars' history, could have been largely destroyed and merged into the global average profile by subsequent mantle dynamics, and it would have left a relatively minor mark on the mantle reference with which later basin-forming impacts are compared; this conclusion does not apply to the crust, however. A more definitive statement would require a combination of a dynamical impact model of the dichotomy-forming event with a long-term evolution model. This would show whether the anomalies of basin-forming impacts of the size considered in our models should rather be compared against a hemispherical reference depending on whether they lie in the Borealis hemisphere or not. If the Borealis-forming impact has left a stronger and more regional signature in the mantle, later impacts would hit a more depleted target than assumed here, and the resulting anomalies would be reduced relative to our predictions.\par
Finally, if the initial state of the mantle is not as well mixed as we assumed, a larger amount of water may be retained in the deeper part of the mantle and increase its conductivity. In such models, the most deep-reaching impact anomalies would probably produce an even stronger contrast and be more visible than in our examples, especially if the mantle has generally become more well-mixed in its evolution after the impact. Indeed, the high conductivity in the wadsleyite stability field of the \citet{CiTa14} model suggests that the water content in the deeper mantle might be substantially higher than in our models; the possibility of such a deep water reservoir and its petrological implications have been studied experimentally by \citet{MeGr06} and \citet{Pomm:etal12}. Therefore we recommend that more effort be put into obtaining electromagnetic sounding data, because of all geophysical methods, this seems to be the one most likely to be able to constrain the open question of the water content of the martian mantle, and in combination with other methods it holds considerable promise for exposing impact signatures beneath uncertain candidate structures and for elucidating how efficient melt extraction in the final stages of crater formation is.

\section{Conclusions}
Combining dynamical mantle convection models and various sets of mineral physics data, we predict the spatial patterns of electrical conductivity, bulk sound speed, and density in the present-day martian mantle. By applying a simplified representation of meteorite impact effects, we also assess the detectability of traces large impacts may have left in the interior of the planet. We find that for a high efficiency of the extraction of melt generated immediately by the shock from the impact, anomalies may be produced whose compositional component survives billions of years and may be detected today. Deep-reaching measurements of the electrical conductivity combined with gravity data seem to be a promising approach for studying the subsurface of martian impact structures. On a more general scale, deep electromagnetic sounding of the mantle also holds some promise to constrain the water content of the martian mantle and the radial distribution of water in it. If the deep high-conductivity layer proposed by \citet{CiTa14} is confirmed, this would suggest limited mixing of the mantle during martian evolution.\par
A dedicated survey targeted at impact basins should combine the powers of multiple geophysical methods and also include orbiting spacecraft that provide gravity data as well as radar data to clarify the electrical conductivity structure of the hydrosphere or cryosphere. The main module would have to be at least one lander in the basin featuring a seismometer, whose principal purpose would be to constrain the thickness of the crust and the velocity structure especially in the upper mantle, and a magnetometer for long-duration magnetic variation recordings that probe the interior down to the CMB. Ideally, a second station at an out-of-basin location like InSight provides data of the same type as a reference; otherwise, such a second station would have to be deployed. Any additional stations forming a regional seismic and electromagnetic network at the basin would obviously help to clarify the three-dimensional subsurface structure of the basin.\par
Given that substantial amounts of melt are not only produced by the shock during the impact but also by convective processes in its aftermath, it will still be challenging to clarify the extent of impact melt extraction even if gravity, seismic, and electromagnetic observations were available. The models do suggest, however, that especially for impacts large enough to affect the mantle, impact melt extraction results in a significantly higher depletion and produces distinct anomalies in all three observables, although their amplitude is not large in all cases, and it would be difficult to distinguish it from other types of heterogeneity on the basis of amplitude alone in a one-dimensional profile. Therefore the combination of several methods is important, and structural information from a small network is probably necessary to come to a decision. If the massive production of melt could be linked with some certainty to an impact as the cause and the extent of its extraction be elucidated, then this could underpin a proposed association of compositional heterogeneity of martian mantle reservoirs that has been deduced from meteorite analyses with very large early impacts in the context of planetary accretion \citep[e.g.,][]{Borg:etal16}.\par
Regardless of whether impact melts are extracted or not, the post-impact dynamics point to substantial magmatic activity after the impact, which should lead to significant crustal rejuvenation. This new crust is unlikely to have the same porosity as the ancient pre-impact megaregolith, and its formation will leave behind an anomalously depleted mantle residue. Even though our model is simplified in some respects and hence probably overestimates post-impact crust production, we recommend to replace the notion of uniform densities for the crust and mantle with models that allow for this degree of heterogeneity in the inversion of gravity data.

\appendix
\section{Electrical conductivity formulas and parameters}\label{app:elpar}
We have compiled and evaluated data for individual mineral conductivities from several studies and derived parameter sets of fits to functional models of the general form of the Arrhenius-type law Eq.~\ref{eq:sig-genArrh}, omitting or fixing certain terms as required by the available data; this is indicated by ``after'' in the corresponding reference in Table~\ref{tab:condpar}. The parameters are for the lattice conductivity of the crystals; grain boundary processes were found to be negligible for the grain sizes assumed here \citep[cf.][]{AGJones16}.\par
\afterpage{%
\begin{landscape}
\begin{longtable}[c]{lccccccll}
\caption{Parameters for Eq.~\ref{eq:sig-genArrh} of the individual mineral phases. If $V$ was determined, $H$ values are activation energies instead of enthalpies (n.d. = not determined; \textsuperscript{*}: diffusion coefficient in m\textsuperscript{2}/s, using Eq.~\ref{eq:sigp}).\label{tab:condpar}}\\\toprule
Mineral&$\sigma_0$&$q$&$H$ or $E$&$\alpha$&$V$&$\beta$&Buffer&Reference\\
&(S/m)&&\multicolumn{2}{c}{\hspace*{2em}(kJ/mol)}&\multicolumn{2}{c}{\hspace*{2em}(cm\textsuperscript{3}/mol)}&&\\\midrule
Olivine\\
\hspace*{3ex}ionic\hspace{1.5em}\begin{minipage}{0.07\textwidth}(Mg)\\(O)\end{minipage}&
\begin{minipage}{0.16\textwidth}\centering 62000\\$2.2\times10^6$\end{minipage}&--&
\begin{minipage}{0.15\textwidth}\centering 208.4\\296.2\end{minipage}&--&
\begin{minipage}{0.12\textwidth}\centering 5.3\\$-1.1$\end{minipage}&--&--&\begin{minipage}{0.3\textwidth}\citet{Yosh:etal17}\\\citet{Yosh:etal17}\end{minipage}\\
\hspace*{3ex}small-polaron&525&1&189.1&143.8&$-0.01$&0.22&MMO&\citet{Yosh:etal12a}\\
\hspace*{3ex}proton\hspace{1.3em}\begin{minipage}{0.07\textwidth}$\parallel \mathrm{a}$\\$\parallel \mathrm{b}$\\$\parallel \mathrm{c}$\end{minipage}&\begin{minipage}{0.15\textwidth}\centering 0.2\textsuperscript{*}\\$10^{-5}$\textsuperscript{*}\\$3.16\times10^{-4}$\textsuperscript{*}\end{minipage}&--&
\begin{minipage}{0.15\textwidth}\centering 229\\172\\188\end{minipage}&--&n.d.&--&&
\begin{minipage}{0.3\textwidth}\citet{Nove:etal17b}\\\citet{Nove:etal17b}\\\citet{Nove:etal17b}\end{minipage}\\
$\beta$-olivine\\
\hspace*{3ex}small-polaron&279.75&1&159.5&123.3&n.d.&n.d.&MMO&after \citet{Yosh:etal12a}\\
\hspace*{3ex}proton&2500&1&80.1&89.6&n.d.&--&MMO&\citet{YoKa12}\\
$\gamma$-olivine\\
\hspace*{3ex}small-polaron&1885&1&186.2&140.9&$-0.33$&0.72&MMO&after \citet{Yosh:etal12a}\\
\hspace*{3ex}proton&97724&0.69&104&--&n.d.&--&MMO&\citet{XHuan:etal05}\\
Orthopyroxene\\
\hspace*{3ex}ionic&855610&--&242.2&--&4.15&--&NNO&\citet{BZhYo16}\\
\hspace*{3ex}small-polaron&163&1&224.8&192&1.06&0.12&NNO&after \citet{BZhYo16}\\
\hspace*{3ex}proton&38019&1&81&35.8&n.d.&--&MMO&\citet{BZhan:etal12}\\
Clinopyroxene\\
\hspace*{3ex}ionic&144.5&--&102&--&n.d.&--&NNO&\citet{XYang:etal11}\\
\hspace*{3ex}proton&660693&1.13&71&--&n.d.&--&NNO&\citet{XYang:etal11}\\
Plagioclase\\
\hspace*{3ex}ionic&13183&--&161&--&n.d.&--&NNO&\citet{XYang:etal12}\\
\hspace*{3ex}proton&14125&0.83&77&--&n.d.&--&NNO&\citet{XYang:etal12}\\
Garnet\\
\hspace*{3ex}ionic&257026&--&272.2&--&n.d.&--&MMO&after \citet{Roma:etal06}, py\textsubscript{100}\\
\hspace*{3ex}small-polaron&7479.5&1&231.8&199.9&2.5&n.d.&MMO&after \citet{Roma:etal06},\\
&&&&&&&&\citet{DaKa09a}\\
\hspace*{3ex}proton&295121&1.09&68&--&n.d.&--&MMO&\citet{KaWa13}\\
Majorite\\
\hspace*{3ex}ionic&3927&--&169.8&--&0&--&MMO&\citet{Yosh:etal08b}\\
\hspace*{3ex}small-polaron&144349&3.794&58.3&$-110$&0&0&MMO&after \citet{Yosh:etal08b}\\
\hspace*{3ex}proton&\multicolumn{8}{c}{assumed same as garnet}\\
Stishovite\\
\hspace*{3ex}proton&104200&1&103.2&85.1&--&--&MMO?&\citet{Yosh:etal14}\\
\bottomrule
\end{longtable}
\end{landscape}}
In the case of olivine, we decided to follow \citet{AGJones16} and calculated the conductivity from hydrogen diffusion data using a combination of the Nernst--Einstein equation and the classic Arrhenius law for lattice diffusion,
\begin{equation}
\sigma_\mathrm{p}=f_\mathrm{H}D_\mathrm{0p}\exp\left(-\frac{E_\mathrm{p}+pV_\mathrm{p}}{RT}\right)\frac{C_\mathrm{H}e^2}{k_\mathrm{B}T},\label{eq:sigp}
\end{equation}
rather than electrical conductivity measurements, but we opted to use the new data from \citet{Nove:etal17b} instead of those from \citet{DuFrTy12} used by \citet{AGJones16}; in this equation, the correlation correction factor $f_\mathrm{H}$ is assumed to be 1, $D_\mathrm{0p}$ is the diffusion pre-factor for protons, and $e$ and $k_\mathrm{B}$ are the elementary charge and Boltzmann's constant, respectively. We replaced the lattice diffusion data used by him with the effective-medium value determined from the new data with the diffusion equivalent of Eq.~\ref{eq:sigEM} and $x_i=1/3$ for each of the principal directions \citep[cf.][]{Stroud75}; we did not use the simplified conductivity expression derived from the geometric mean offered by \citet{Nove:etal17b} themselves.\par
The experimental data must be rescaled to the oxygen fugacity conditions assumed for the martian mantle, which are considered to lie close to the iron--wüstite buffer \citep[e.g.][]{Herd03}. \citet{KaWa13} discussed the role of oxygen fugacity for the conductivity terms related to iron and water and proposed that $\sigma_\mathrm{sp}$ and $\sigma_\mathrm{p}$ need to be rescaled with a factor $(f_{\mathrm{O}_2}/f_{\mathrm{O}_2,0})^s$, where $f_{\mathrm{O}_2,0}$ is the oxygen fugacity at which the experimental data were obtained; for those cases for which the relation was investigated, it was found that $s=0.17$ for dry conditions and $s=-0.1$ for wet conditions. Considering the substantial uncertainty, an empirical formula for the oxygen fugacity should suffice to calculate the ratio, and so we use the model function
\begin{equation}
\lg f_{\mathrm{O}_2}=b+\frac{c(p-10^5)-a}{T}\label{eq:lgfO2}
\end{equation}
with the fitting parameters $a$, $b$, and $c$ as given by \citet{LDai:etal11} to calculate the oxygen fugacities for the iron--wüstite (IW), Ni--NiO (NNO), and Mo--\chem{MoO}{2} (MMO) buffers. The oxygen fugacity factor is not applied to the calculation for proton conduction in olivine, because there are too few constraints on the influence of the oxygen fugacity on diffusion.

\section{Sensitivity considerations}\label{app:sens}
\subsection{Mineral conductivity}
Eq.~\ref{eq:sig-genArrh} quantifies how the electrical conductivity of a specific mineral depends on temperature, pressure, and composition, and hence the derivatives with respect to these variables offer some insight into its sensitivity to them. It is therefore worthwhile to consider some of them in detail.\par
For the sensitivity to temperature and pressure variations, it should suffice to consider the simpler classic Arrhenius form, i.e., Eq.~\ref{eq:sig-genArrh} with $q_i=\alpha_i=\beta_i=0$. Then the $T$ derivative takes the form
\begin{equation}
\partdrv{\sigma}{T}=
\sigma_0\frac{E+pV}{RT^2}\exp\left(-\frac{E+pV}{RT}\right)=
\frac{E+pV}{RT^2}\sigma.
\end{equation}
With regard to sensitivity, the prefactor in the second forms of these equations, i.e., the relative change $\partial\ln\sigma/\partial T$, is the term of interest. For typical activation enthalpies (cf. Table~\ref{tab:condpar}), it ranges from several per cent to a few tens of per cent per kelvin at low $T$ and drops to values below 1\%/K at temperatures expected in the lower part of the lithosphere, i.e., the temperature sensitivity is highest at low $T$.\par
For the typical activation volumes of a few cm\textsuperscript{3}/mol, the pressure dependence is always weak compared to the temperature dependence of the derivative. The pressure derivative of the simplified Eq.~\ref{eq:sig-genArrh} is
\begin{equation}
\partdrv{\sigma}{p}=-\sigma_0\frac{V}{RT}\exp\left(-\frac{E+pV}{RT}\right)=
-\frac{V}{RT}\sigma.
\end{equation}
The relative change $\partial\ln\sigma/\partial p$ is only $T$-dependent and is on the order of 0.1/GPa or less for typical activation volumes even at low temperatures, and hence $\sigma$ generally shows quite little variation with pressure. Note that the activation volume can be positive or negative.\par
For the compositional (iron or water content) derivatives, we consider the general form of Eq.~\ref{eq:sig-genArrh},
\begin{subequations}
\begin{equation}
\sigma=\sigma_0 C^q \exp\left(-\frac{E-\alpha C^{\frac{1}{3}}+p(V-\beta C)}{RT}\right);
\end{equation}
for the proton conductivity, we either have $q=1$, $\beta=0$ or replace $q\neq 1$ and set $\alpha=\beta=0$. The $C$ derivative of the general form is
\begin{equation}
\partdrv{\sigma}{C}=
\left(\frac{q}{C}+\frac{\alpha C^{-\frac{2}{3}}+3p\beta}{3RT}\right)\sigma.
\end{equation}
\end{subequations}
The sensitivity to compositional changes depends itself on composition and possibly on temperature; there is also a potential, albeit negligible, $p$ dependence. For the sensitivity to iron content, the range of interest is $0.5\lesssim \mathrm{Mg\#}\lesssim 0.98$ for planetary materials. Although the sensitivity is largest at low $T$ and Mg\#, it is still very significant at more common iron contents and mantle temperatures; for instance, under mid-mantle conditions at the martian reference value of $\mathrm{Mg\#}=0.75$, $\partial\ln\sigma/\partial C_\mathrm{Fe}$ is on the order of $-0.1$/Mg\#. For the sensitivity to water content, the dependence on water content is most important, and it is greatest at low concentrations. At the water concentrations of a few tens of ppm expected for the martian mantle, $\partial\ln\sigma/\partial C_\mathrm{w}$ is on the order a few per cent per ppm water.\par
As the total conductivity of a phase is the sum of all charge carrier contributions, so is the derivative with respect to any of the variables, but typically, one contribution dominates under any given conditions. If the rock has undergone melting, its Mg\# will increase and its water content drop, potentially entailing a shift in the dominant charge carrier and the sensitivity to further changes of conditions.

\subsection{Composite (rock) conductivity}
The picture is complicated further when considering rock composed of different materials, and closed-form solutions become too unwieldy to be useful. Some insight can nonetheless be gained from inspection of the derivatives of the self-consistent effective medium value Eq.~\ref{eq:sigEM}. Denoting the derivative with respect to some variable by a prime, we have
\begin{equation}
0=\sum\limits_{i=1}^N \left(x'_i\frac{\sigma_i-\sigma_\mathrm{EM}}{\sigma_i+2\sigma_\mathrm{EM}}\right.\\
{}+\left. x_i\frac{(\sigma'_i-\sigma'_\mathrm{EM})(\sigma_i+2\sigma_\mathrm{EM})-(\sigma_i-\sigma_\mathrm{EM})(\sigma'_i+2\sigma'_\mathrm{EM})}{(\sigma_i+2\sigma_\mathrm{EM})^2}\right).
\end{equation}
The $\sigma_i$ in this equation are the sums over all charge carriers of the derivative with respect to the variable of interest outlined above, for the $i$th mineral constituent. The crucial detail to notice here, however, is that it is not only derivatives of conductivities that appear here but also derivatives of volume fractions, $x'_i$, which consist basically of two types of contributions.\par
The first type is related to the thermoelastic properties of the individual mineral phases in the assemblage. If sensitivity to temperature variations is considered, the ratios of the thermal expansion coefficients of the different minerals in the assemblage always contribute to $x'_i$, but as the expansivity for all common minerals is usually of the order $10^{-5}$/K, this influence is probably negligible in general. In a similar vein, the ratios of compressibilities of the different mineral phases contribute to $x'_i$ in the case of pressure variations, but again, this contribution is probably minor, at least for small $p$ variations; if larger $p$ ranges are considered, the changes in the mineral proportions that reflect the general change in the petrological composition of the rock along the areotherm will be more important. Changes in iron or water content affect the molar volume of mineral solid solutions and hence also have an impact on $x'_i$, because these changes will not be the same for all mineral phases in an assemblage. However, these changes are again minor and may be neglected.\par
The second type of contributions to $x'_i$ is related to the mineralogical composition of the rock, i.e., to how the mass proportions of the mineral phases change in response to variations in the thermodynamic conditions and bulk chemical composition of the rock; these changes in mass fractions translate into changes in the volume fractions $x_i$. The changes related to this second type are defined by the phase diagram of the rock and are strongest at phase boundaries or when the rock melts. Especially in the case of melting they are expected to be important, as the concomitant shifts in volumetric proportions are much larger than those of the first type and can fundamentally change the relative contributions of phases of strongly contrasting conductivities to the bulk $\sigma$.

\subsection{Seismic velocity and density}\label{app:sens-seis}
The sensitivity of density to temperature changes follows directly from the definition of the thermal expansivity
\begin{equation}
\alpha_V=\frac{1}{V}\partdrv{V}{T}=-\frac{1}{\varrho}\partdrv{\varrho}{T},
\end{equation}
which for mantle minerals is typically on the order of $10^{-5}$/K. In the case of $v_\mathrm{B}=\sqrt{K_S/\varrho}$, the temperature derivative can be written as
\begin{equation}
\partdrv{v_\mathrm{B}}{T}=\frac{v_\mathrm{B}}{2}\left(\partdrv{\ln K_S}{T}+\alpha_V\right).
\end{equation}
The reason for the reduction with $T$ is then that $\partial\ln K_S/\partial T$ is negative and its absolute value is about an order of magnitude larger than $\alpha_V$ for many mantle minerals \citep[e.g.,][]{DuAn89}. By contrast, $\varrho$ and $\sigma$ are reduced along more depleted areotherms, whereas $v_\mathrm{B}$ is increased.

\subsection{Application to model areotherms}\label{sect:sens-areo}
\begin{figure*}
\centering
\includegraphics[width=\textwidth]{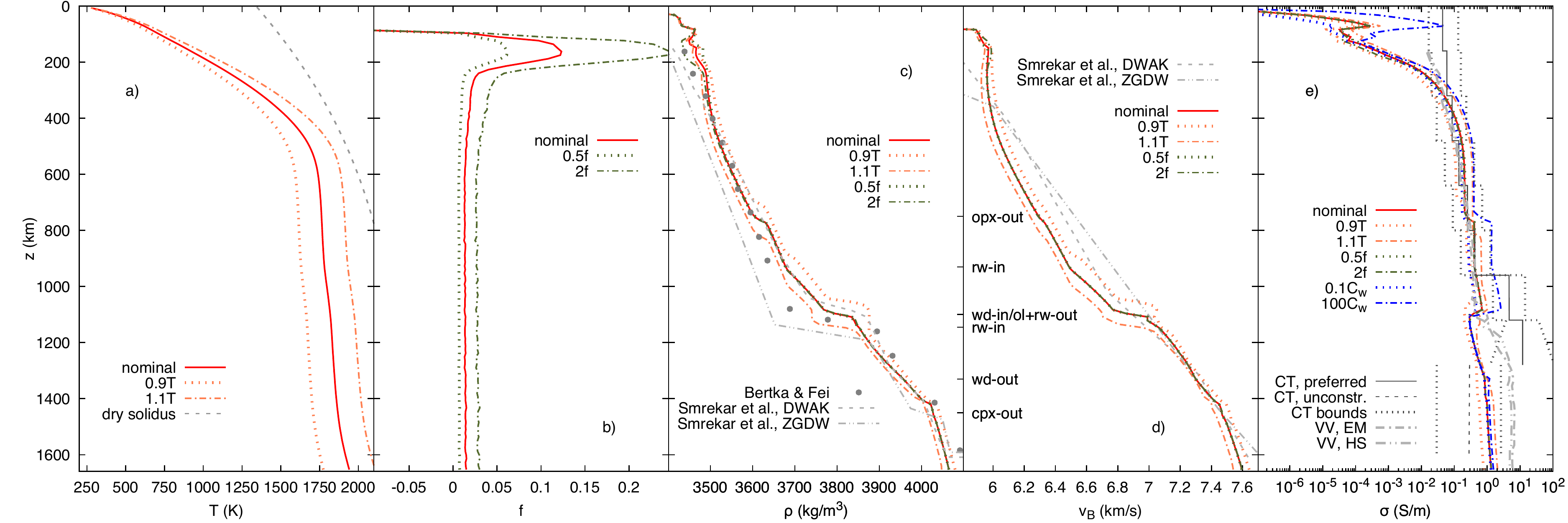}
\caption{Sensitivity of observables to variations in temperature, melting degree, and water content along the global mean areotherm of the impact-free model with 36\,ppm initial bulk water over the full depth range of the mantle. Also shown are the same experimental, inversion, and observation values as in Fig.~\ref{fig:zref} as well as the position of phase transitions for the nominal areotherm. In the density (c) and bulk sound speed (d) profiles, the modified water contents are not plotted, because no dependence of $\varrho$ and $v_\mathrm{B}$ on water was included in the underlying database; it is considered to be negligible at the low concentrations of interest here, as explained in the main text. The sensitivity to variations in $f$ is visibly different from the nominal profile only in the uppermost mantle, where the most depleted material is located. Only the $\sigma(z)$ profiles show different variation patterns for all three variables $T$, $f$, and $C_\mathrm{w}$, whereby the sensitivity to $f$ variations is quite small and the differences are most pronounced in the uppermost mantle and the crust; this applies also to the sensitivity plots for the uppermost 520\,km in Figs.~S11 and S12. For most of the depth range constrained by measurements, the model profiles from \citet{VeVa16} fall within the range covered by our sensitivity tests.\label{fig:sens-glob}}
\end{figure*}

In order to assess the sensitivity of the electrical conductivity $\sigma$ as well as of the density $\varrho$ and the bulk sound speed $v_\mathrm{B}$ to changes in temperature $T$, melting degree $f$, and water content $C_\mathrm{w}$ in the context of the bulk material of our models, we used the models with 36\,ppm initial bulk water content as a starting point. We took the global mean areotherm from the model without impacts (cf. Fig.~\ref{fig:zref}, red curves) and the in-crater averaged areotherms from the other models (cf. Fig.~\ref{fig:zoomHU}, red and orange curves) and derived from them three pairs of modified depth profiles for which observables were calculated; in each pair, either $T$, $f$, or $C_\mathrm{w}$ were varied while keeping everything else at the nominal value. For the temperature pair, the temperature along the areotherm was recalculated as $T\leftarrow T_\mathrm{min}+c(T-T_\mathrm{min})$ for $c=0.9$ or 1.1. For the melting degree pair, where $f$ is effectively a summary measure of iron content and mineral proportion variation, $f$ is scaled by a factor 0.5 or 2, respectively; note that for this particular calculation, water concentration is decoupled from $f$ and kept at the nominal values. The scaling is only applied to the mantle part of the depth profile, which is of more interest to us, because we do not have sufficient information on the dependence of mineralogy and composition of crustal material on the melting degree of its source rock to implement it in similar form as for mantle material. For the water content pair, the water concentration was scaled by a factor 0.1 or 100, respectively, but without taking into account that the solubility limit would likely be exceeded in the crust and give rise to hydrous minerals; the effect of $C_\mathrm{w}$ is neglected in the calculation of $\varrho$ and $v_\mathrm{B}$. The results are shown in Figs.~\ref{fig:sens-glob}, S11, and S12.\par
As these figures show, $\varrho$ and $v_\mathrm{B}$ are reduced along hotter areotherms, whereas $\sigma$ is increased along them. There is no significant difference in sensitivity between models with impact melt retention and with complete extraction (Figs.~S11 and S12). Comparison of the experimental densities determined for a martian model composition by \citet{BeFe98} with the different areotherms (Fig.~\ref{fig:sens-glob}) shows that slightly higher mantle temperatures would give an even better agreement, especially in the interval between $\sim$750\,km and wd-in/ol+rw-out. They would also slightly improve the match with the MGS-derived conductivity profile by \citet{CiTa14} in the mid-mantle and with the position of the olivine--wadsleyite transition determined from autocorrelation reflectivity analysis of InSight seismological data by \citet{DeLe20}. However, neither the $T$ nor the $C_\mathrm{w}$ increases we tried would quite suffice to reproduce the high conductivity in the crust and the jump to conductivities of several S/m at around 1000\,km depth \citet{CiTa14} deduce from their inversion, although especially the increased water content would help to come much closer. From the geodynamical point of view, the dilemma with a potentially slightly higher temperature is that it would result in an even thicker crust \citep[cf.][]{Rued:etal13a}.\par
The profiles indicate that a 10\% uncertainty in the temperature would translate into a moderate signal in all three observables $\varrho$, $v_\mathrm{B}$, and $\sigma$ at all depths and would probably be large enough to be resolved. By contrast, a factor of 2 uncertainty in the melting degree $f$ does not produce a significant variation unless the reference $f$ exceeds $\sim 0.05$, whereby $v_\mathrm{B}$ appears to be particularly insensitive. The steep decline of $\sigma$ in the uppermost mantle is chiefly controlled by temperature, but the $100C_\mathrm{w}$ curves suggest that it would be modulated significantly in case of higher water concentrations; in the models with a bulk initial water content of 36\,ppm, water concentrations in the deeper mantle have dropped to a few ppm during evolution, and so the $100C_\mathrm{w}$ curves would correspond to a concentration of a few hundred ppm, which is at the upper end of the range of estimates from recent years. The generally high conductivities and the depth of the conductivity minimum in the lithospheric mantle should then be useful indicators of the water content of the mantle, because these features could not easily be explained by other factors.

\section*{CRediT authorship contribution statement}
Thomas Ruedas: Conceptualization, Funding acquisition, Investigation, Software, Writing -- original draft, Writing -- review \& editing.
Doris Breuer: Writing -- review \& editing

\section*{Acknowledgments}
We thank Kai Wünnemann and Lukas Manske for instructive discussions about issues related to coupled melt and impact dynamics and for sharing some of their simulation results before publication, and Sebastiano Padovan, Daniel Wahl, Olivier Verhoeven, and Harro Schmeling for helpful advice concerning crustal density structure, gravity anomalies, electrical conductivity measurements, and porous flow estimates respectively. Comments by an anonymous referee on an earlier version of the manuscript and by two anonymous referees on this version helped us to clarify and improve various aspects of this paper. We also thank Wyatt Du Frane and Davide Novella for helpful explanations on H diffusion and its implications for conductivity in olivine reported in \citet{Nove:etal17b}. TR was supported by DFG (Deutsche Forschungsgemeinschaft) grant Ru 1839/2-1. DB was supported by DFG~-- Project-ID 263649064~-- TRR 170. This is TRR~170 publication no.~120. The numerical calculations were carried out on the supercomputer ForHLR~II funded by the Ministry of Science, Research and the Arts Baden-Württemberg and by the Federal Ministry of Education and Research of Germany.

\end{document}